\documentclass{article}

\usepackage[title]{appendix}
\PassOptionsToPackage{numbers, compress}{natbib}

\usepackage[nonatbib, preprint]{neurips_2025}

\usepackage{cite}  

\usepackage{adjustbox}
\usepackage{amsmath}

\usepackage[utf8]{inputenc} 
\usepackage[T1]{fontenc}    
\usepackage{hyperref}       
\usepackage{url}            
\usepackage{booktabs}       
\usepackage{amsfonts}       
\usepackage{nicefrac}       
\usepackage{microtype}      
\usepackage{xcolor}         
\usepackage{tablefootnote}
\usepackage{graphicx} 
\usepackage{wrapfig}
\title{Moonbeam: A MIDI Foundation Model Using Both Absolute and Relative Music Attributes}

\author{%
  Zixun Guo \\
Centre for Digital Music\\
  Queen Mary University of London\\
  \texttt{zixun.guo@qmul.ac.uk} \\
  \And
  Simon Dixon \\
  Centre for Digital Music \\
 Queen Mary University of London \\
 \texttt{s.e.dixon@qmul.ac.uk}
}

\begin{document}

\newcommand{\modelname}{\texttt{Moonbeam}}

\newcommand{\totaltraininghours}{81.6K}
\newcommand{\newattention}{Multidimensional Relative Attention}
\newcommand{\newattentionshort}{MRA}
\newcommand{\smallmodelsize}{309M}
\newcommand{\largemodelsize}{839M}
\newcommand{\TODO}[1]{\textcolor{red}{[TODO: #1]}}

\maketitle

\begin{abstract}
\modelname{} is a transformer-based foundation model for symbolic music, pretrained on a large and diverse collection of MIDI data totaling \totaltraininghours{} hours of music and 18 billion tokens. 
\modelname{} incorporates music-domain inductive biases by capturing both absolute and relative musical attributes through the introduction of a novel domain-knowledge-inspired tokenization method and \newattention{} (\newattentionshort{}), which captures relative music information without additional trainable parameters. Leveraging the pretrained \modelname{}, we propose 2 finetuning architectures with full anticipatory capabilities, targeting 2 categories of downstream tasks: symbolic music understanding and conditional music generation (including music infilling). Our model outperforms other large-scale pretrained music models in most cases in terms of accuracy and F1 score across 3 downstream music classification tasks on 4 datasets. Moreover, our finetuned conditional music generation model outperforms a strong transformer baseline with a REMI-like tokenizer. We open-source the code, pretrained model, and generated samples on \href{https://github.com/guozixunnicolas/Moonbeam-MIDI-Foundation-Model}{Github}.
\end{abstract}

\section{Introduction} \label{sec:intro}

In recent years, AI foundation models have demonstrated effectiveness in many domains. These models are usually pretrained using a large amount of openly available data and then finetuned for many downstream tasks. In the field of music, there has been progress in symbolic music pretraining, but most models are trained on data in ABC format \cite{Yuan_chatmusician, qu2025mupt}, which represents non-expressive music scores, or using BERT-like \cite{Devlin_bert} architectures \cite{wu_clamp, wu_clamp2, musicbert}, which limits their generative capabilities. Since MIDI is able to represent expressively performed music as well as music scores, a large amount of symbolic music is stored in MIDI format, yet remains underutilized in large-scale pretraining. 

The main challenge in training large symbolic music models lies in the tokenizer, which converts symbolic music into tokens. Existing tokenizers often focus on tokenizing a specific data type, thus limiting their generalizability and hindering future finetuning. Some tokenizers \cite{remi, cp_transformer, wu_clamp, wu_clamp2, musicbert, Yuan_chatmusician, qu2025mupt, popmag, multitrack_music_transformer} are only able to tokenize music scores where time is represented in terms of bars and beats, so that they are unable to model expressive music, while others \cite{remi, pianist8, this_time_with_feeling, musictransformer} are limited to tokenizing single-instrument data (e.g., piano data) only. Moreover, tokenizing multi-track symbolic music often results in excessively long input sequences, which either require substantial computational resources \cite{anticipatory_music_transformer} or rely on techniques such as byte-pair-encoding \cite{symphonynet, bpe_music} to reduce computation costs. Meanwhile, relative musical attributes play a crucial role in our perception of music \cite{musictransformer, FME, lattner_interval_2018, lattner_interval_2018_predictive}. It is often possible to change absolute attributes (e.g., the dynamics or key) while still maintaining the identity of the music via the relative relationships. Although relative music attributes have proven to be important music domain inductive biases in small-scale models, as including them consistently results in superior model performance \cite{Liutkus_rel_attn, musictransformer, relative_attention, FME}, more recent large symbolic music models \cite{wu_clamp, wu_clamp2, qu2025mupt, anticipatory_music_transformer} often rely on a standard transformer decoder, discarding these important inductive biases, potentially due to the lack of scalable and effective model architectures.





To address these, we present \modelname{}, a transformer-based foundational model for symbolic music. It leverages a novel, efficient, and generalizable tokenizer and \newattention{} (\newattentionshort{}), a novel attention mechanism proposed to incorporate music-domain inductive biases. More specifically, our tokenizer is able to tokenize a wide range of MIDI data, including performance and score data, single and multi-instrument data, while maintaining compact sequence lengths. Moreover, \newattentionshort{} encodes both absolute and relative positional information in a multi-dimensional space. In contrast, in RoPE \cite{rope}, relative positional information is only calculated in a 1-dimensional space (i.e., token indices). In our case, each music token resides in a 5-dimensional space, representing onset time, duration, octave, pitch class, and velocity. \newattentionshort{} enables the model to explicitly encode both absolute and relative music information for all 5 attributes separately.


\modelname{} has been pretrained on \totaltraininghours{} hours (around 18 billion tokens) of freely available MIDI data, spanning across different instruments (e.g., guitar, piano, bass, saxophone, drums, multitrack), formats (e.g., performance and score MIDI), and genres (e.g., jazz, rock, pop, classical). Table \ref{tab:training_datasets} in Appendix \ref{sec:training_datasets} lists a summary of the training data used. \modelname{} was trained at two scales: a small model with \textasciitilde \smallmodelsize{} parameters, and a medium-sized model with \textasciitilde \largemodelsize{} parameters. To the best of our knowledge, \modelname{} is the first autoregressive transformer-based foundation model that is trained at scale on a large and diverse collection of MIDI data. 

\modelname{} can serve as a versatile backbone for downstream tasks, and we demonstrate this by finetuning it on two tasks: symbolic music understanding and controllable music generation (including music infilling). In the music understanding tasks, including player, emotion and composer classification across 4 datasets, our finetuned model outperforms other state-of-the-art (SOTA) large-scale pretrained symbolic music models \cite{wu_clamp2, musicbert} in most cases. Moreover, we propose a unified finetuning framework that leverages our pretrained model for controllable music generation and music infilling, without using an additional encoder \cite{guo_musiac} or an interleaved control sequence \cite{anticipatory_music_transformer}. Since note onset times are represented using absolute values in our data representation, our proposed method has full anticipatory ability \cite{anticipatory_music_transformer}, enabling its music infilling capability. Our proposed method outperforms a transformer baseline \cite{lee_commu} with a REMI-like \cite{remi} tokenizer.

The contributions of the paper are summarized as follows: 1. To the best of our knowledge, \modelname{} is the first large symbolic music model pretrained on a large and diverse corpus of MIDI data; 2. We propose a novel and efficient tokenization method that supports conversion of a wide range of MIDI data, including performance, score, single- and multi-instrument MIDI. Our tokenization method also demonstrates strong generalization capabilities, allowing it to handle previously unseen data during finetuning; 3. We propose \newattention{} (\newattentionshort{}), which encodes relative positional information in more than one dimension. This enables \modelname{} to leverage music-specific inductive biases by incorporating both absolute and relative music information; 4. Leveraging the pretrained \modelname{}, we present two finetuning architectures for downstream tasks, including symbolic music understanding, conditional music generation and music infilling. In most cases, our finetuned models outperform other transformer baselines.

\section{Related Work} \label{sec:lit_review}
We provide a brief overview of popular tokenization methods for symbolic music. The \emph{MIDI-like} tokenizer \cite{this_time_with_feeling} and the \emph{REMI} \cite{remi} tokenizer focus on tokenizing piano MIDI. In the \emph{MIDI-like} tokenizer, time is represented using \texttt{<note\_on>}, \texttt{<note\_off>}  and \texttt{<timeshift>} tokens, while \emph{REMI} uses \texttt{<bar>}, \texttt{<beat>} and \texttt{<duration>} tokens to represent time. In \emph{REMI}, bar and beat information are derived from external beat and downbeat trackers as they are not directly encoded in performance MIDI. In \cite{bpe_music, impact_time_duration}, \texttt{<bar>} and \texttt{<beat>} tokens are replaced by \texttt{<timeshift>} tokens to avoid the use of external beat and downbeat trackers. To tokenize multi-instrument MIDI, \emph{REMI+} \cite{figaro} extends \emph{REMI} by adding instrument and time signature tokens. This unavoidably increases the sequence length, leading to higher computational costs. To mitigate this, \textit{PopMAG} \cite{popmag} groups each triplet of pitch, duration, and velocity events into a single compound token, while \cite{bpe_music} employs byte pair encoding (BPE) to learn event groupings dynamically. In parallel, another line of research focuses on combining various attributes of MIDI events into compound tokens. This has proven to effectively reduce the sequence length and, meanwhile, increase the inference efficiency \cite{cp_transformer, nested_music_transformer, multitrack_music_transformer, symphonynet}. More recently, we have seen an increasing trend to represent symbolic music in ABC format as text \cite{qu2025mupt, Yuan_chatmusician}, modeled directly with LLMs. 

The transformer backbones used in symbolic music modeling can be broadly classified into two categories: the standard transformer decoder \cite{transformer, llama, radford2019language_gpt2} and its variants (e.g., Transformer-XL \cite{transformer-xl} and linear transformer \cite{linear_transformer}), as used in \cite{remi, cp_transformer, anticipatory_music_transformer, figaro, multitrack_music_transformer, bpe_music}; and the transformer decoder with relative attention \cite{relative_attention, musictransformer, FME, Liutkus_rel_attn}.
As computational complexity in the transformer increases quadratically with input sequence length, Transformer-XL \cite{remi, popmag, lakhnes} and linear transformers \cite{cp_transformer, symphonynet} have been adopted to increase computational efficiency. Another line of research emphasizes the inclusion of relative music information during the attention calculation. In \cite{Liutkus_rel_attn}, the use of relative positional embeddings in the transformer model results in more coherent music generation compared to using only absolute positional embeddings. Music Transformer \cite{musictransformer} uses relative attention \cite{relative_attention} to capture latent relationships between inputs at all timesteps. The RIPO transformer \cite{FME} extends this by explicitly incorporating relative index, pitch and onset during the attention calculation. 

Several works focus on large-scale pretraining for symbolic music. MusicBERT\cite{musicbert} introduces a bar-level masking technique for BERT-like \cite{Devlin_bert} pretraining. Clamp \cite{wu_clamp, wu_clamp2} converts symbolic music into text using direct type casting followed by a novel patching technique to reduce the sequence length. Subsequently, the music encoder in Clamp is pretrained using a bidirectional masked language model. MusicBERT and Clamp both adopt a BERT-like \cite{Devlin_bert} architecture, which limits their generative capabilities. In parallel, MIDI-GPT \cite{Pasquier_midigpt} and Anticipatory Music Transformer \cite{anticipatory_music_transformer} are trained on GigaMIDI \cite{Lee_gigamidi} and LakhMIDI \cite{raffel_lakh} respectively, using autoregressive transformer decoders.  More recently, LLMs have been leveraged \cite{Yuan_chatmusician, qu2025mupt} to conduct pretraining and finetuning using symbolic music in ABC format. However, since ABC data encodes music scores rather than expressive music performances, these models are unable to take advantage of the large quantity of performance data that exists in formats such as MIDI, and that can be generated by automatic transcription from audio.


\section{Method} \label{sec:method}
\begin{figure}
    \centering
    \includegraphics[width=\linewidth]{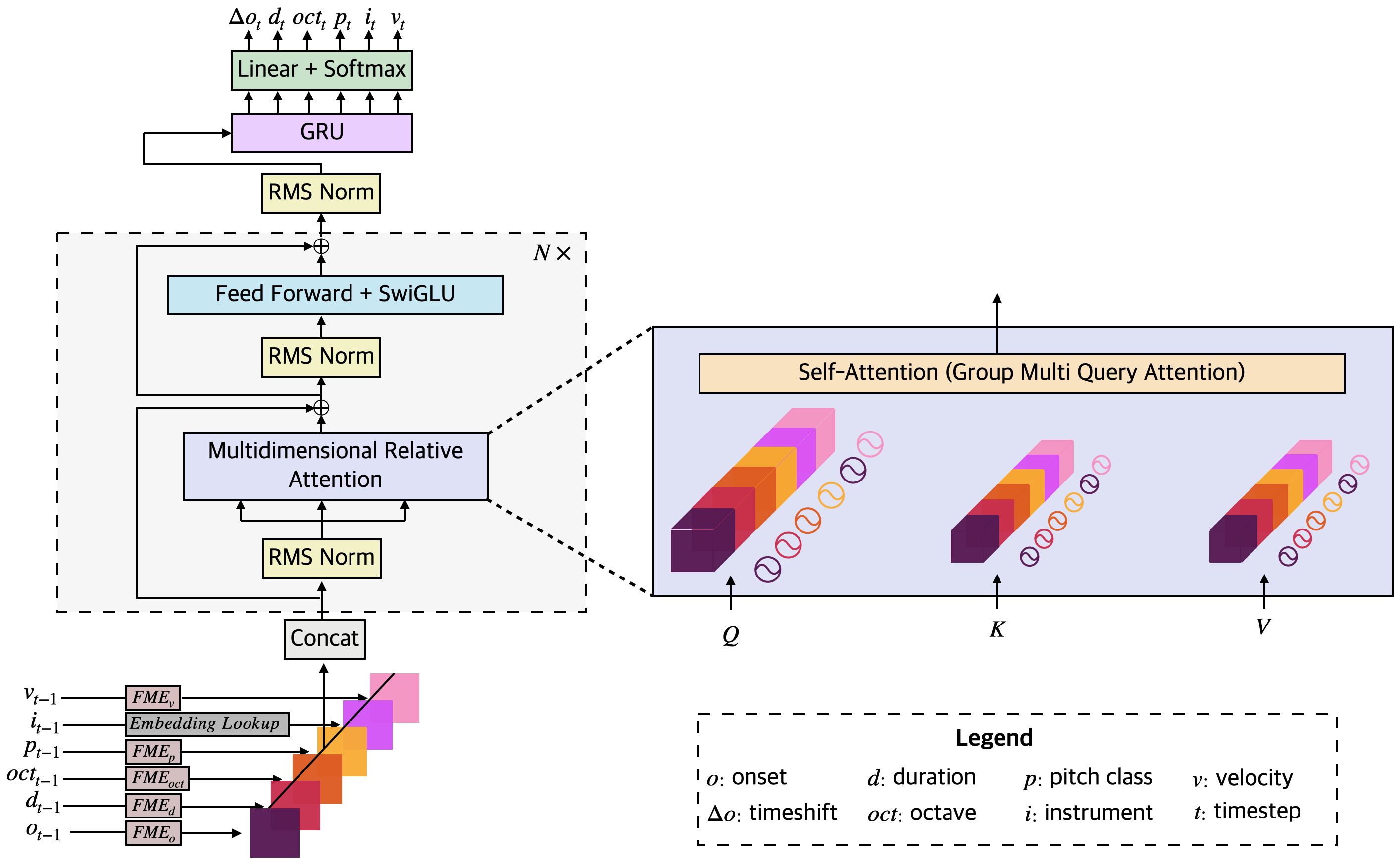}
    \caption{\modelname{} Model Architecture.}
    \label{fig:model}
\end{figure}


The overall model architecture and tokenization procedure are illustrated in Figure~\ref{fig:model}. \modelname{} models the following distribution: $p_{\theta}(\tilde{x}_t|x_1, ..., x_{t-1})$ 
 using a neural network parametrized by $\theta$. $x$ is a compound token $(o, d, oct, p, i, v)$ that represents the onset, duration, octave, pitch class, instrument and velocity of a musical event. Given the past $t-1$ tokens, \modelname{} autoregressively predicts the next event $\tilde{x} = (\Delta{o}, d, oct, p, i, v)$ at time step $t$, where $\Delta{o}$ represents delta onset or timeshift. Note that we input absolute onsets $o$ instead of timeshifts $\Delta{o}$ into the model, as transformer models have been shown to struggle with arithmetic \cite{bradshaw2025ariamidi}, and the model would otherwise need to aggregate all previous timeshifts to compute the next onset time.  Directly inputting the absolute onset effectively mitigates this \cite{bradshaw2025ariamidi}. This also directly benefits our downstream music infilling task, since using absolute onsets facilitates the alignment between the generated and control sequences. We then use a standard cross entropy loss function to model the categorical distribution of each sub-event of $\tilde{x}$.
 
 Next, we begin by introducing our tokenization method, followed by our model architecture, which incorporates our proposed \newattention{} (\newattentionshort{}) and a GRU sub-decoder. Finally, we propose two finetuning architectures for two downstream tasks: symbolic music understanding and conditional music generation.

\subsection{Tokenization}\label{sec:method_tokenization}

%

As a foundation model, \modelname{} is intended for finetuning or adaptation to downstream tasks or new data. Hence, we aim to design a generalizable tokenizer capable of tokenizing a wide range of MIDI data, not limited to a specific type (e.g., single-instrument, score MIDI). Meanwhile, the tokenizer should also incorporate music domain knowledge. To achieve this, we extend the tokenization method introduced in \cite{FME}, which is limited to monophonic, single-instrument MIDI. We use a mixture of standard embedding and Fundamental Music Embedding (FME) \cite{FME}, a music-domain-knowledge inspired embedding function, to tokenize and embed each attribute of $x$ then concatenate them together. As a result, our tokenizer can handle multi-instrument, polyphonic, and expressive MIDI data. More specifically, among these attributes: $o, d, oct, p, i, v$ in $x$, only the instrument $i$ is well-suited for encoding using standard embeddings, while the other attributes contain relative music information and should be translation-equivariant (or translation-invariant, depending on tasks) in the embedding space. FME is a suitable embedding function as it preserves the relative music information in the embedding space 
\cite{FME, mustango}. Following \cite{FME}, for all non-music tokens (e.g., \texttt{<start-of-sequence>}, \texttt{<end-of-sequence>}, \texttt{<classification>}), we use standard embeddings. 
The overall tokenization process is illustrated in Figure \ref{fig:model} (below the transformer layers). 

There are several advantages of the proposed tokenization method: First, it demonstrates strong generalization and extensibility since it can tokenize a wide range of MIDI data (e.g., multi-track, performance and score MIDI). Moreover, because FME is a continuous sinusoidal embedding function, it can handle interpolated and extrapolated inputs (e.g., microtonal inputs, pitch bends, longer note duration), leaving room for future finetuning without requiring additional trainable parameters. In contrast, standard tokenization methods require the creation of additional entries in the embedding lookup table to process unseen inputs. Second, the proposed tokenizer preserves relative music information in the embedding space, which is an important inductive bias to the model as mentioned in Section~\ref{sec:intro}. Third, note onset times are represented using their absolute value, unlike in other tokenization methods, which typically rely on timeshifts or relative timing aligned to a metrical grid \cite{bpe_music, impact_time_duration}. FME enables the conversion of absolute onsets $o$ into meaningful latent representations without significantly increasing the input dictionary size, as would occur with standard embedding lookups. As a result, \modelname{} does not have to spend extra effort learning to add all the timeshifts to represent onsets. More importantly, representing onsets using absolute values facilitates temporal alignment during future finetuning, if other control sequences (e.g., a second MIDI control sequence, sequential data from other modalities such as video) are introduced. Last but not least, the proposed tokenizer requires fewer trainable parameters. Standard embedding lookup tables require $\mathit{dictionary\_size} \times \mathit{hidden\_size}$ trainable parameters, whereas the majority of trainable parameters in our proposed tokenization method come from the linear layer in FME \cite{FME}, totaling approximately $\mathit{hidden\_size}^2$. Since the dictionary size required is typically much larger than $\mathit{hidden\_size}$, our method significantly decreases the number of trainable parameters in the tokenizer. Nevertheless, it outperforms the standard tokenization method, as will be shown in Section \ref{sec:results}.



\subsection{Model Architecture}

There are two core differences between the transformer used in LLaMa \cite{llama} and \modelname{}: 1. We replace the multi-headed group query attention \cite{Ainslie_GQA} with our proposed \newattention{}(\newattentionshort{}), with which the model can leverage both absolute and relative music information without additional trainable parameters; 2. We add a GRU decoder to sequentially decode each sub-event of the next event $\tilde{x}$, from the transformer output, similar to \cite{nested_music_transformer}.




\subsubsection{\newattention{} (\newattentionshort{})}
Our proposed \newattentionshort{} is inspired by the Rotary Position Embedding (RoPE) \cite{rope}, which encodes relative positional information in a 1-dimensional (1-D) space as shown in Equations \ref{eq:rope_q} to \ref{eq:rope_qk}. A query vector $\mathbf{Q}^m = \mathbf{W}_q \mathbf{x}_m$ at position $m$ and a key vector $\mathbf{K}^n$ at position $n$, are rotated by $m\theta$ and $n\theta$ respectively, where $\theta$ is a predefined set of constants, and $\mathbf{W}_q$ and $\mathbf{W}_k$ are trainable linear layers for query and key respectively. During the attention calculation, as shown in Equation \ref{eq:rope_qk}, their relative position $m-n$ is encoded without additional trainable parameters, differently from \cite{relative_attention}.

\begin{equation} \label{eq:rope_q}
    f_q(\mathbf{x}_m, m) = (\mathbf{W}_q \mathbf{x}_m)e^{i m \theta} = \mathbf{Q}^m e^{i m \theta}
\end{equation}
\begin{equation}\label{eq:rope_k}
    f_k(\mathbf{x}_n, n) = (\mathbf{W}_k \mathbf{x}_n)e^{i n \theta}= \mathbf{K}^n e^{i n \theta}
\end{equation}
\begin{equation}\label{eq:rope_qk}
\langle \mathbf{Q}^m e^{i m \theta}, \mathbf{K}^n e^{i n \theta} \rangle = \Re\left[(\mathbf{Q}^m e^{i m \theta})(\mathbf{K}^n e^{i n \theta})^*\right] = \Re\left[\mathbf{Q}^m \mathbf{K}^{n^*} e^{i (m-n) \theta}\right]
\end{equation}

In our proposed \newattentionshort{}, we extend RoPE to encode relative positional information beyond 1-D. We first define the position of an event at position index $t$ in the event sequence as $v(t) = (v_1(t), ... , v_N(t))$, where $N$ denotes the number of intrinsic dimensions and $v_g$ represents orthogonal decompostion of the position along the $g$-th dimension. In the 1-D case where $N=1$ and $v(t)=t$, all attention heads are rotated by $t \theta$ as in Equation \ref{eq:rope_q} and \ref{eq:rope_k}. To extend this into higher dimensions while still preserving relative positional information along each axis, we partition all attention heads into groups and rotate each group using orthogonal decompositions of $v$. More specifically,  given the input tensor $\mathbf{X} \in \mathbb{R}^{l \times d}$, where $l$ is the sequence length and $d$ is the hidden dimension, we first compute the query $\mathbf{Q} =  \mathbf{W}_q \mathbf{X}$ and key $\mathbf{K} = \mathbf{W}_k \mathbf{X} $ vectors. Following MHA \cite{transformer}, we partition $\mathbf{Q}$ and $\mathbf{K}$ along the feature dimension into $H$ heads, each with dimensionality $d_h = d / H$ as follows: $\mathbf{Q} = [\mathbf{Q}_1, \mathbf{Q}_2, \dots, \mathbf{Q}_H], \mathbf{K} = [\mathbf{K}_1, \mathbf{K}_2, \dots, \mathbf{K}_H], \mathbf{Q}_h, \mathbf{K}_h \in \mathbb{R}^{l \times d_h}$. Then, we assign the $H$ heads into $G$ groups, where $G \geq N$ is required to ensure that each intrinsic dimension can be distinctly associated with at least one group of attention heads. We then assign each group $\mathcal{G}_g$ with its corresponding position decomposition $v_g(t)$ along each orthogonal direction. Here, $\mathcal{G}_g \subseteq \{1, 2, \dots, H\}$ represent the set of heads in group $g$, such that: $\bigcup_{g=1}^G \mathcal{G}_g = \{1, 2, \dots, H\}, \mathcal{G}_g \cap \mathcal{G}_{g'} = \emptyset \text{ for } g \neq g'.$



The rotation is applied as: $f_q^g(\mathbf{Q}, t_q) = \mathbf{Q}_{\mathcal{G}_g} e^{i v_g(t_q) \theta_g}, f_k^g(\mathbf{K}, t_k) = \mathbf{K}_{\mathcal{G}_g} e^{i v_g(t_k) \theta_g}$
where $\mathbf{Q_{\mathcal{G}_g}}$ and $\mathbf{K_{\mathcal{G}_g}}$ are concatenated query and key vectors for all heads in group $\mathcal{G}_g$, $\theta_g$ is a set of predefined constants for group $g$, and $t_g$ and $t_k$ are positional indices for query and key vectors. Subsequently, the rotated tensors across all groups are concatenated: $f_q(\mathbf{Q}, t_q) = \text{Concat}\left(f_q^1(\mathbf{Q}, t_q), \dots, f_q^G(\mathbf{Q}, t_q)\right)$, $
f_k(\mathbf{K}, t_k) = \text{Concat}\left(f_k^1(\mathbf{K}, t), \dots, f_k^G(\mathbf{K}, t_k)\right)
$. Finally, we follow the standard attention calculation \cite{rope} and multiply $f_q(\mathbf{Q}, t_q)$ and $f_k(\mathbf{K}, t_k)$ to obtain the final attention score. We provide a full derivation of how relative positional information is encoded in higher dimensions in Appendix \ref{sec:MRA_derivation}. Compared to other additive relative attention mechanisms \cite{relative_attention, musictransformer, FME}, \newattentionshort{} is more parameter-efficient, as it does not require additional trainable parameters.

In our specific case for music, each event $x$ resides in a 5-dimensional (5-D) space ($N=5$) with each dimension representing onset, duration, octave, pitch class and velocity, respectively. We set $G=6$ (the extra head group comes from the instrument attribute $i$) and we set $v_g(t)$ as below. With \newattentionshort{}, \modelname{} can encode both absolute and relative music information within the 5-D space for all music attributes. Our approach is inherently different from existing transformer architectures for music \cite{mmm, anticipatory_music_transformer, figaro, multitrack_music_transformer, bpe_music, remi, cp_transformer} where each token’s position is defined by its position index in the input sequence, which lacks clear musical meaning. 

\[
v_g(t) = 
\begin{cases} 
o_t, & \text{if } g = 1 \text{ or } g = 5, \\
d_t, & \text{if } g = 2, \\
oct_t, & \text{if } g = 3, \\
p_t, & \text{if } g = 4, \\
v_t, & \text{if } g = 6.
\end{cases}
\]
\subsubsection{GRU Decoder and the Training Objective}
At each time step, the transformer outputs a single tensor, which represents a latent representation for generating the next event at time step $t$. To decode all attribute tokens, this tensor is used as the initial hidden state of a GRU, which then sequentially decodes all 6 attributes: $ \left( \Delta{o}_t, d_t, oct_t, p_t, i_t, v_t \right)$. The main reason for decoding the tokens sequentially is to model the conditional distribution of subsequent attribute tokens based on previously sampled tokens, as these tokens are correlated. For instance, if a sub-event $p_t$ with a high pitch is sampled, it is unlikely that the instrument $i_t$ producing this pitch is a bass. Since only six attribute tokens need to be sampled (sequence length = 6) per timestep, a GRU is sufficient and we choose not to use another transformer due to its complexity. During training, we use a standard cross entropy loss to model the categorical distribution of each sub-event in $\tilde{x}$.


\begin{figure}
    \centering
    \includegraphics[width=\linewidth]{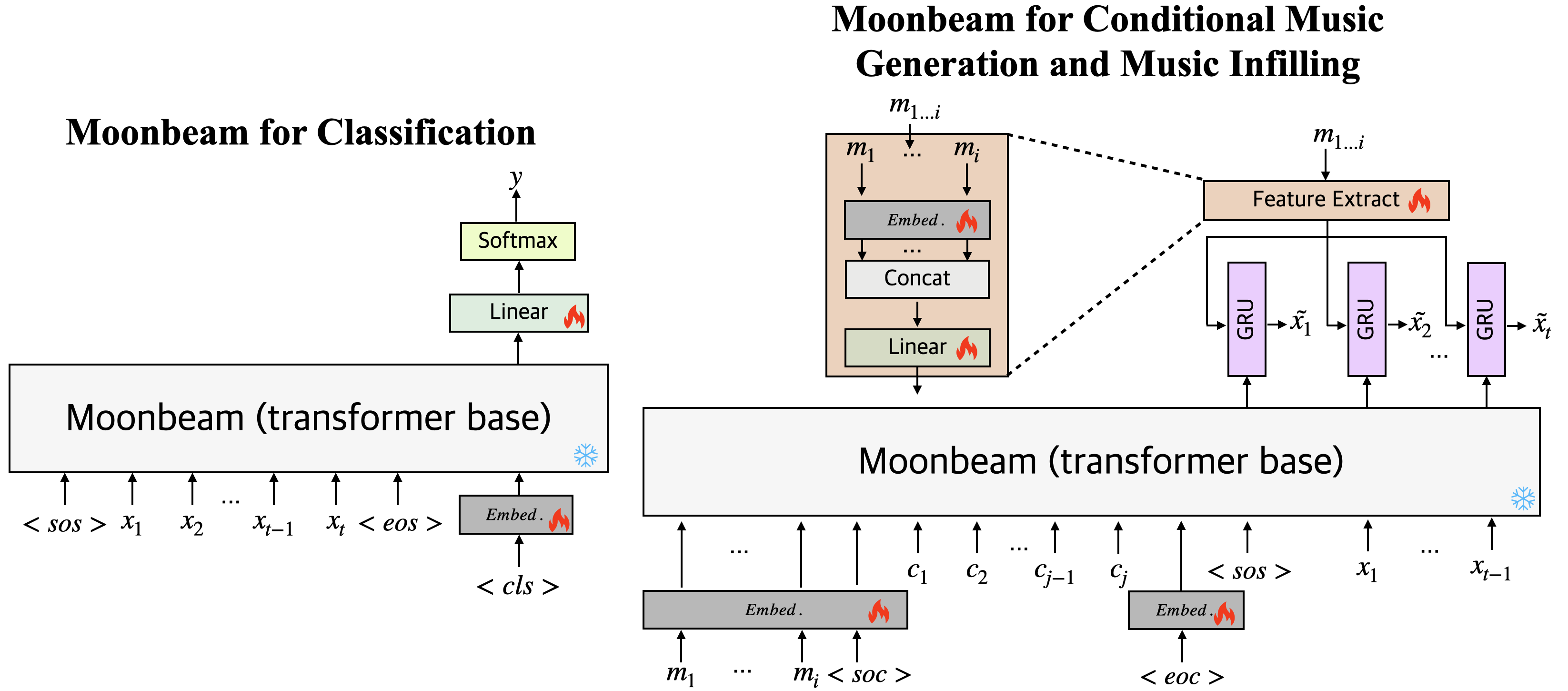}
    \caption{\modelname{} Finetuning Architecture.}
    \label{fig:model_finetune}
\end{figure}

\subsection{Finetuning \modelname{} for Music Classification and Conditional Generation} \label{sec:method_finetune}

In the finetuning classification architecture shown in Figure \ref{fig:model_finetune}, we append one classification token \texttt{<cls>} after each input sequence and replace the GRU decoder with a linear layer for classification. During finetuning, we apply LoRA \cite{hu_lora} while keeping the embedding layer for the \texttt{<cls>} token and the linear layer for classification fully trainable. We use a cross-entropy loss to optimize the model.

Conditional music generation is a task in which a model learns to compose music based on specified input conditions. We define $i$ input conditions $m_i$ that have no temporal dimension (e.g., genre, key). 
One sub-task within this broader definition is music infilling \cite{anticipatory_music_transformer,guo_musiac}, where the model learns to generate the missing parts or ``fills'' $x$, given an incomplete composition $c_1, ... c_{j}$, with $j$ events. 


The proposed unified framework, as illustrated in Figure \ref{fig:model_finetune}, aims to generate music $x$ autoregressively based on both non-temporal conditions $m$ and the music infilling condition $c$. During training, the input conditions $m$ and $c$ are prepended to the input sequence $x$. We separate condition signals from each other and from music events using a \texttt{<soc>} and \texttt{<eoc>} token. We use an additional fully trainable embedding layer to embed $m$ and the 2 separation tokens. Since the music infilling condition $c$ shares the same format as the music events $x$, they can be input directly to \modelname{}'s transformer base. The non-temporal conditions $m$ are also added to the GRU decoder of \modelname{} via a feature extractor that consists of a trainable embedding layer and a linear layer. The unified framework has full anticipatory capability \cite{anticipatory_music_transformer}, since note onsets are in absolute time and due to the use of \newattentionshort{}. When generating the event $x_t$ at timestep $t$, the model can attend to previously generated tokens and the entire control sequence $c$, including both past and future control events. Since $x$ and $c$ are both represented in absolute time, with \newattentionshort{}, the model calculates the relative onsets between $x_t$ and all events in $c$, thereby allowing the model to determine $x_t$'s relative position within the control sequence.

\section{Experimental Settings} \label{sec:experiment_setting}


%
\subsection{Pretraining Settings and Ablation Studies} \label{sec:pretraining}

We train \modelname{} at different scales: the model configurations and the data preprocessing pipeline are listed in Table \ref{tab:model_size} in Appendix \ref{sec:model_configs} and Appendix \ref{sec:data_preprocessing} respectively. To increase training efficiency, we adopt LLaMa's \cite{llama} data concatenation technique during data loading. Specifically, we concatenate the input sequence until the fixed sequence length ($L=1024$) and use a block-diagonal causal attention mask to ensure data from different input samples will not attend to each other during training.  We use an Adam optimizer with an initial learning rate of $3\times 10^{-4}$ and decay factor of $0.85$ per training epoch (due to the large size of our training data, the training does not exceed 9 epochs). We employ the Distributed Data Parallel (DDP) strategy to distribute training across two A100 GPUs \footnote{When training \modelname{} (S), we used 2 A100s with 40GB of RAM; and 2 A100s with 80GB RAM when training \modelname{} (M)}. Additionally, we leverage mixed precision training \cite{Micikevicius_mixed_precision_training} to improve efficiency and reduce memory usage. Training \modelname{} (S) and \modelname{} (M) takes 54 hours and about 15 days, respectively.

We perform a series of ablation studies using \modelname{} (S) to verify the effectiveness of each module in \modelname{} and compare each model's test perplexity using one randomly selected 5\% test set of the Lakh Dataset \cite{raffel_lakh}. In each baseline model, we replace one module of \modelname{} at a time and monitor the test perplexity. Specifically, we substitute FME \cite{FME} with a standard embedding lookup (except for onsets: $o$, which would otherwise result in an excessively large dictionary size). We also replace \newattentionshort{} with either a standard multi-head group-query attention mechanism \cite{llama}, or a variant of \newattentionshort{}. In the variant of \newattentionshort{}, we rotate all attention heads with the sum of all music attribute values. This variant also encodes the overall relative music information (see Appendix \ref{sec:MRA_variant} for proof). Finally, we replace the GRU decoder with a 2-layer MLP with a hidden size of 1360 to model the unconditional distribution of each sub-event in $\tilde{x}$, as defined in Section \ref{sec:method}. We ensure the baseline model with replaced modules has approximately the same number of trainable parameters as the original model.

\subsection{Downstream Task Settings}


In the first set of music classification tasks, we compare our model with other large-scale pretrained music models: M3 \cite{wu_clamp2}, Clamp2 \cite{wu_clamp2}, and MusicBERT \cite{musicbert}, evaluating them across a broader set of datasets and tasks, including player classification using PiJAMA30 \cite{pijama} and Pianist8 \cite{pianist8}, emotion classification using Emopia \cite{hung_emopia} and composer classification using a self-constructed dataset: Giant-Piano-MIDI 30 (GPM30) derived from GPM \cite{kong_giantpianomidi} to address the the lack of a high-quality composer classification dataset (see Appendix \ref{sec:finetuning_settings} for details). We include the full finetuning settings as well as a preliminary test to validate our finetuning architecture in Appendix \ref{sec:finetuning_settings}. To ensure fairness, we compare different models at the piece-level rather than clip-level, since the data representation varies from model to model. During evaluation, we observe that some baseline models cannot process long inputs, for instance the maximum input length for MusicBERT \cite{musicbert} is 1024.  Following \cite{kong_giantpianomidi}, for each model, when the input exceeds the maximum sequence length allowed, we use a non-overlapping sliding window to perform inference on different segments of the piece and average the logits to select the most likely output. We use accuracy and F1 score to evaluate these models. 


For the conditional music generation and infilling tasks, we utilize the CoMMU dataset \cite{lee_commu}, composed and performed by professional composers, where each snippet of music $x$ has a paired chordal accompaniment $c$ (converted from chord symbols provided in the dataset), and 12 metadata conditions $m$ (e.g., pitch and velocity range, number of measures, track role). The finetuning settings are provided in Appendix \ref{sec:finetuning_settings}. To validate our finetuning architecture, we conduct a series of ablation tests where we add (or remove) the condition signals $m$ and $c$ at the GRU and (or) the transformer and monitor the test perplexity of each model. Since test perplexity alone is not sufficiently informative, we assess some aspects of the model’s condition-following capability (controllability) by computing three metrics, which can be measured objectively. Specifically, we sample music from these models following the condition in the test set, using greedy sampling to minimize the randomness introduced by distribution-based sampling. The objective controllability metrics are described as follows: 1. the ratio of notes within the correct pitch and velocity range, with tolerances up to $\pm 5$ bins; 
and 2. whether the model could learn when to end a piece given the accompaniment and the duration specified in the metadata. To reflect this, we calculate the differences between the end times of the accompaniment and the generated music and calculate their mean and standard deviation. Note that these objective controllability metrics do not reflect musicality. For example, consider a scenario where a single note is repeatedly played at the correct pitch and velocity for the entire duration of the piece. While this may yield perfect objective scores, the resulting output would be completely non-musical.


After the ablation test, we compare our finetuned model against a strong baseline: a transformer-based model \cite{lee_commu} with a REMI-like \cite{remi} tokenizer, trained with the same data. Based on manual inspection of the generated music, we used top-p sampling with $p = 0.6$ and temperature $t = 0.7$ to balance novelty and coherence for our model, following \cite{anticipatory_music_transformer}. For the baseline model, we adopt the optimal sampling configuration reported in \cite{lee_commu}. We compare these models based on their controllability and musicality. While some controllability metrics can be measured objectively (i.e., aforementioned pitch and velocity accuracy and end-time prediction), other aspects require subjective evaluation. Since the subjective evaluation involves assessing whether the generated music adheres to the input chord and metadata conditions, the listening test requires expertise and can only be carried out by experts. We recruited 20 music experts to participate in our study, of whom 55\% had over 11 years of musical training and 90\% had more than 4 years of training. The music experts are instructed to rate 10 randomly-sampled pairs of music generated from the two models, using the same sets of conditions from the test set. We asked the participants to rate the generated music on a 5-point Likert scale, using the following questions: 1. How well does the generated music fit with the chord condition and 2. the metadata condition?; 3. Does the generated music develop randomly or coherently?; 4. Overall, how much do you enjoy the generated music? 

\section{Results} \label{sec:results}
\subsection{Ablation Studies on \modelname{} (S)}
The results of the ablation studies are shown in Table \ref{tab:ablation}. Our proposed model, \modelname{}, equipped with the proposed tokenizer, \newattentionshort{}, and a GRU decoder, achieves the lowest test perplexity among all models. Notably, replacing the GRU decoder with an MLP results in increased test perplexity, which confirms the interdependencies between each sub-event. Moreover, without incorporating FME or \newattentionshort{}, the two ablated models rely mostly on absolute music information and lack relative context. This also results in higher test perplexity. These findings highlight the importance of incorporating music-domain-specific inductive biases into the transformer architecture, leading to enhanced model performance. Finally, the model with our proposed \newattentionshort{} outperforms its variant, demonstrating the effectiveness of \newattentionshort{}.



\begin{table}
  \caption{Ablation study on \modelname{} (S)}
  \label{tab:ablation}
  \centering
  \resizebox{0.8\columnwidth}{!}{%
    \begin{tabular}{lcccc}
      \toprule
      Model & Tokenizer Emb. Func. & Attn. Type & Sub-decoder & Test PPL \\
      \midrule
      \modelname{} w/o FME & Standard Emb & \newattentionshort{} & GRU & 4.216 \\
      \modelname{} attn. variant & FME & \newattentionshort{} rotate all heads & GRU & 2.512 \\
      \modelname{} standard attn. & FME & Standard attention & GRU & 2.512 \\
      \modelname{} w/o GRU & FME & \newattentionshort{} & MLP & 3.245 \\
      \midrule
      \modelname{} & FME & \newattentionshort{} & GRU & \textbf{2.423} \\
      \bottomrule
    \end{tabular}%
  }
\end{table}

\subsection{Downstream Music Classification Results}
Table \ref{tab:downstream_classification} presents the results of downstream music classification tasks. Across most tasks, \modelname{} (M) consistently achieves the highest performance. Despite its smaller size and being pretrained with fewer data, our small model: \modelname{} (S) performs reasonably well and even outperforms all other large models on PiJAMA30. For Pianist8, \modelname{} (M) achieves high accuracy and F1 score, likely due to the dataset featuring pianists from very different genres, making them easily distinguishable. On the Emopia dataset, while M3 achieves the best performance, 
\modelname{} (M) performs competitively, within about 2\% of M3. 
Finally, on GPM30, \modelname{} (M) again leads on accuracy and F1. These results show the robustness and effectiveness of \modelname{} in various music classification tasks, highlighting its potential for broader applications. 


\begin{table}
  \caption{Downstream music classification results}
  \label{tab:downstream_classification}
  \centering
    \resizebox{0.95\columnwidth}{!}{%
  \begin{tabular}{lcccccccccc}
    \toprule
    Dataset     & \multicolumn{2}{c}{PiJAMA30\cite{pijama}} & \multicolumn{2}{c}{Pianist8\cite{pianist8}}   & \multicolumn{2}{c}{Emopia\cite{hung_emopia}}    & \multicolumn{2}{c}{GPM30}    \\
    Task        & \multicolumn{2}{c}{Pianist Classification} & \multicolumn{2}{c}{Pianist Classification}   & \multicolumn{2}{c}{Emotion Classification}    & \multicolumn{2}{c}{Composer Classification}    \\
    \cmidrule(r){2-3} \cmidrule(r){4-5} \cmidrule(r){6-7} \cmidrule(r){8-9}
    Model          & Acc         & F1 Macro   & Acc         & F1 Macro   & Acc         & F1 Macro   & Acc         & F1 Macro   \\
    \midrule
    Clamp 2 \cite{wu_clamp2}   & 0.440        & 0.219      & 0.892        & 0.891      & 0.659       & 0.652      & 0.644       & 0.549      \\
    M3 \cite{wu_clamp2}       & 0.262       & 0.077      & 0.784       & 0.786      & \textbf{0.715}       & \textbf{0.688}       & 0.385        & 0.160      \\
    MusicBert \cite{musicbert} & 0.550       & 0.452      & 0.811       & 0.806      & 0.682       & 0.674      & 0.630        & 0.575      \\
\midrule
    \modelname{} (S)  & 0.649       & 0.596      & 0.811      & 0.804      & 0.636       & 0.623     & 0.541      & 0.470    \\
    \modelname{} (M)  & \textbf{0.679}       & \textbf{0.638}      & \textbf{0.946}      & \textbf{0.947}      & 0.693       & 0.682     & \textbf{0.648}       & \textbf{0.635}    \\
    \bottomrule
  \end{tabular}}
\end{table}

\subsection{Downstream Conditional Music Generation Ablation Studies and Evaluation}
From the results of the ablation studies in Table \ref{tab:downstream_gen_obj}, we observe firstly that the metadata conditioning alone is insufficient for the model to end a piece at the correct time, as indicated by the large mean and standard deviation of timing difference. Moreover, including the metadata condition at the GRU is essential for velocity and pitch accuracy. 
Second, when we only add the chord condition, the model fails to generate notes in the correct pitch and velocity ranges, as expected. However, it does learn where to end a piece, possibly based on the 
relative position of generated notes within the overall chord sequence. Third, allowing a tolerance up to $\pm 5$ bins greatly improves results, showing that many of the errors are small. These observations validate our finetuning architecture - by incorporating the metadata condition into both the GRU and the Transformer, and applying the chord condition only to the Transformer, the model achieves the lowest test perplexity and learns to generate music with appropriate pitch, dynamics and duration. We also compared these objective metrics with the baseline model and our model slightly underperforms the baseline model in terms of pitch and velocity accuracy. The listening test results in Table \ref{tab:subjective_test}, however, show that \modelname{} was rated significantly better than the baseline model on all aspects of the evaluation, indicating that it can generate music with greater musicality and better adherence to metadata conditions. 

\begin{table}
  \caption{Ablation studies and objective evaluation for downstream conditional music generation}
  \label{tab:downstream_gen_obj}
  \centering
    \resizebox{0.95\columnwidth}{!}{%
  \begin{tabular}{llcccccccccc}
    \toprule
    Metadata Location & Chord Location & Test PPL & \multicolumn{4}{c}{Velocity Accuracy} & \multicolumn{4}{c}{Pitch Accuracy} & Timing \\
    \cmidrule(r){4-7} \cmidrule(r){8-11}
     & & & tol=0 & tol=1 & tol=3 & tol=5 & tol=0 & tol=1 & tol=3 & tol=5 & Mean / STD \\

    \midrule
    GRU & None & 2.367 & 0.879 & 0.916 & 0.952 & 0.971 & 0.701 & 0.775 & 0.844 & 0.916 & -0.898 / 6.025 \\
    Transformer & None & 2.356 & 0.358 & 0.423 & 0.494 & 0.531 & 0.254 & 0.275 & 0.342 & 0.393 & 0.177 / 2.726 \\
    GRU+Transformer & None & 2.344 & 0.872 & 0.906 & 0.954 & 0.964 & 0.870 & 0.915 & 0.941 & 0.971 & 5.205 / 21.751 \\
     
    \midrule
    None & GRU & 2.412 & 0.208 & 0.232 & 0.286 & 0.325 & 0.157 & 0.169 & 0.239 & 0.250 & -2.177 / 6.687 \\
    None & Transformer & 2.295 & 0.213 & 0.269 & 0.309 & 0.326 & 0.105 & 0.115 & 0.163 & 0.225 & -0.151 / 1.367 \\
    None & GRU+Transformer & 2.301 & 0.215 & 0.260 & 0.290 & 0.305 & 0.113 & 0.121 & 0.139 & 0.187 & 0.102 / 0.601 \\

    \midrule
    GRU+Transformer & GRU+Transformer & 2.266 & 0.811 & 0.871 & 0.905 & 0.958 & 0.828 & 0.887 & 0.940 & 0.958 & -0.112 / 1.008 \\
    GRU+Transformer & Transformer & \textbf{2.254} & 0.862 & 0.895 & 0.932 & 0.968 & 0.851 & 0.889 & 0.934 & 0.952 & 0.233 / 3.397 \\
    \midrule
    Baseline \cite{lee_commu} & Baseline \cite{lee_commu} & N/A & 0.997 & 0.999 & 1.000 & 1.000 & 0.915 & 0.957 & 0.988 & 0.993 & -0.887 / 1.530 \\
    \bottomrule
  \end{tabular}}
\end{table}

\begin{table}
  \caption{ Human evaluation of paired comparisons of music generated by different systems following the same condition. p-values are reported using a Wilcoxon signed rank test.}
  \label{tab:subjective_test}
  \centering
  \resizebox{0.75\columnwidth}{!}{%
    \begin{tabular}{lcccc}
      \toprule
      Model & Chord-Generation-Fit & Metadata-Generation-Fit & Coherence & Overall \\
      \midrule
      \modelname{} & \textbf{3.955 $\pm$ 0.968} & \textbf{3.950 $\pm$ 0.934} & \textbf{3.940 $\pm$ 0.965} & \textbf{3.600 $\pm$ 1.047} \\
      Transformer Baseline \cite{lee_commu} & 3.210 $\pm$ 1.054 & 3.215 $\pm$ 1.129 & 3.105 $\pm$ 1.029 & 2.885 $\pm$ 1.131 \\
      \midrule
      p-value & 1.424e-14 & 1.618e-12 & 1.590e-17 & 9.678e-12 \\
      \bottomrule
    \end{tabular}
  }
\end{table}

\section{Conclusion}\label{sec:conclusion}
We present \modelname{}: a transformer-based MIDI foundation model pretrained using a large and diverse collection of MIDI data. Using both absolute and relative music attributes through a novel tokenizer and our 
\newattention{}, \modelname{} outperforms other baselines on a broad range of tasks and datasets. Our two finetuning architectures for downstream classification and conditional generation tasks were shown to be effective in multiple experiments. We envision \modelname{} to empower a wide range of downstream tasks in the Music Information Retrieval (MIR) field. Our proposed model architecture has potential applications beyond music, in domains involving multi-dimensional data where relative positional information is crucial, such as robotics. 
\bibliographystyle{IEEEtran}
\bibliography{references}

\begin{thebibliography}{10}
\providecommand{\url}[1]{#1}
\csname url@samestyle\endcsname
\providecommand{\newblock}{\relax}
\providecommand{\bibinfo}[2]{#2}
\providecommand{\BIBentrySTDinterwordspacing}{\spaceskip=0pt\relax}
\providecommand{\BIBentryALTinterwordstretchfactor}{4}
\providecommand{\BIBentryALTinterwordspacing}{\spaceskip=\fontdimen2\font plus
\BIBentryALTinterwordstretchfactor\fontdimen3\font minus \fontdimen4\font\relax}
\providecommand{\BIBforeignlanguage}[2]{{%
\expandafter\ifx\csname l@#1\endcsname\relax
\typeout{** WARNING: IEEEtran.bst: No hyphenation pattern has been}%
\typeout{** loaded for the language `#1'. Using the pattern for}%
\typeout{** the default language instead.}%
\else
\language=\csname l@#1\endcsname
\fi
#2}}
\providecommand{\BIBdecl}{\relax}
\BIBdecl

\bibitem{Yuan_chatmusician}
R.~Yuan, H.~Lin, Y.~Wang, Z.~Tian, S.~Wu, T.~Shen, G.~Zhang, Y.~Wu, C.~Liu, Z.~Zhou, L.~Xue, Z.~Ma, Q.~Liu, T.~Zheng, Y.~Li, Y.~Ma, Y.~Liang, X.~Chi, R.~Liu, Z.~Wang, C.~Lin, Q.~Liu, T.~Jiang, W.~Huang, W.~Chen, J.~Fu, E.~Benetos, G.~Xia, R.~B. Dannenberg, W.~Xue, S.~Kang, and Y.~Guo, ``Chatmusician: Understanding and generating music intrinsically with {LLM},'' in \emph{Findings of the Association for Computational Linguistics, {ACL} 2024, Bangkok, Thailand and virtual meeting, August 11-16, 2024}.\hskip 1em plus 0.5em minus 0.4em\relax Association for Computational Linguistics, 2024, pp. 6252--6271.

\bibitem{qu2025mupt}
X.~Qu, Y.~Bai, Y.~Ma, Z.~Zhou, K.~M. Lo, J.~Liu, R.~Yuan, L.~Min, X.~Liu, T.~Zhang, X.~Du, S.~Guo, Y.~Liang, Y.~Li, S.~Wu, J.~Zhou, T.~Zheng, Z.~Ma, F.~Han, W.~Xue, G.~Xia, E.~Benetos, X.~Yue, C.~Lin, X.~Tan, S.~W. Huang, W.~Chen, J.~Fu, and G.~Zhang, ``Mupt: {A} generative symbolic music pretrained transformer,'' in \emph{The Thirteenth International Conference on Learning Representations, {ICLR} 2025, Singapore, April 24-28}, 2025.

\bibitem{Devlin_bert}
J.~Devlin, M.~Chang, K.~Lee, and K.~Toutanova, ``{BERT:} pre-training of deep bidirectional transformers for language understanding,'' in \emph{Proceedings of the 2019 Conference of the North American Chapter of the Association for Computational Linguistics: Human Language Technologies, {NAACL-HLT} 2019, Minneapolis, MN, USA, June 2-7, 2019, Volume 1 (Long and Short Papers)}.\hskip 1em plus 0.5em minus 0.4em\relax Association for Computational Linguistics, 2019, pp. 4171--4186.

\bibitem{wu_clamp}
S.~Wu, D.~Yu, X.~Tan, and M.~Sun, ``Clamp: Contrastive language-music pre-training for cross-modal symbolic music information retrieval,'' in \emph{Proceedings of the 24th International Society for Music Information Retrieval Conference, {ISMIR} 2023, Milan, Italy, November 5-9}, 2023, pp. 157--165.

\bibitem{wu_clamp2}
S.~Wu, Y.~Wang, R.~Yuan, G.~Zhancheng, X.~Tan, G.~Zhang, M.~Zhou, J.~Chen, X.~Mu, Y.~Gao, Y.~Dong, J.~Liu, X.~Li, F.~Yu, and M.~Sun, ``{CL}a{MP} 2: Multimodal music information retrieval across 101 languages using large language models,'' in \emph{Findings of the Association for Computational Linguistics: NAACL 2025}.\hskip 1em plus 0.5em minus 0.4em\relax Albuquerque, New Mexico: Association for Computational Linguistics, Apr. 2025, pp. 435--451.

\bibitem{musicbert}
M.~Zeng, X.~Tan, R.~Wang, Z.~Ju, T.~Qin, and T.~Liu, ``Musicbert: Symbolic music understanding with large-scale pre-training,'' in \emph{Findings of the Association for Computational Linguistics: {ACL/IJCNLP} 2021, Online Event, August 1-6, 2021}, ser. Findings of {ACL}, vol. {ACL/IJCNLP} 2021.\hskip 1em plus 0.5em minus 0.4em\relax Association for Computational Linguistics, 2021, pp. 791--800.

\bibitem{remi}
Y.-S. Huang and Y.-H. Yang, ``Pop music transformer: Beat-based modeling and generation of expressive pop piano compositions,'' in \emph{Proceedings of the 28th ACM International Conference on Multimedia}, 2020, pp. 1180--1188.

\bibitem{cp_transformer}
W.~Hsiao, J.~Liu, Y.~Yeh, and Y.~Yang, ``Compound word transformer: Learning to compose full-song music over dynamic directed hypergraphs,'' in \emph{Thirty-Fifth {AAAI} Conference on Artificial Intelligence, {AAAI} 2021, Virtual Event, February 2-9}, 2021, pp. 178--186.

\bibitem{popmag}
Y.~Ren, J.~He, X.~Tan, T.~Qin, Z.~Zhao, and T.~Liu, ``Popmag: Pop music accompaniment generation,'' in \emph{{MM} '20: The 28th {ACM} International Conference on Multimedia, Virtual Event / Seattle, WA, USA, October 12-16, 2020}.\hskip 1em plus 0.5em minus 0.4em\relax {ACM}, 2020, pp. 1198--1206.

\bibitem{multitrack_music_transformer}
H.~Dong, K.~Chen, S.~Dubnov, J.~J. McAuley, and T.~Berg{-}Kirkpatrick, ``Multitrack music transformer,'' in \emph{{IEEE} International Conference on Acoustics, Speech and Signal Processing {ICASSP} 2023, Rhodes Island, Greece, June 4-10}, 2023, pp. 1--5.

\bibitem{pianist8}
Y.~Chou, I.~Chen, J.~Ching, C.~Chang, and Y.~Yang, ``{MidiBERT-Piano: Large-scale Pre-training for Symbolic Music Classification Tasks},'' \emph{Journal of Creative Music Systems}, vol.~8, no.~1, 2024.

\bibitem{this_time_with_feeling}
S.~Oore, I.~Simon, S.~Dieleman, D.~Eck, and K.~Simonyan, ``This time with feeling: learning expressive musical performance,'' \emph{Neural Comput. Appl.}, vol.~32, no.~4, pp. 955--967, 2020.

\bibitem{musictransformer}
C.~A. Huang, A.~Vaswani, J.~Uszkoreit, I.~Simon, C.~Hawthorne, N.~Shazeer, A.~M. Dai, M.~D. Hoffman, M.~Dinculescu, and D.~Eck, ``Music transformer: Generating music with long-term structure,'' in \emph{7th International Conference on Learning Representations, {ICLR} 2019, New Orleans, LA, USA, May 6-9}, 2019.

\bibitem{anticipatory_music_transformer}
J.~Thickstun, D.~L.~W. Hall, C.~Donahue, and P.~Liang, ``Anticipatory music transformer,'' \emph{Trans. Mach. Learn. Res.}, vol. 2024, 2024.

\bibitem{symphonynet}
J.~Liu, Y.~Dong, Z.~Cheng, X.~Zhang, X.~Li, F.~Yu, and M.~Sun, ``Symphony generation with permutation invariant language model,'' in \emph{Proceedings of the 23rd International Society for Music Information Retrieval Conference, {ISMIR} 2022, Bengaluru, India, December 4-8}, 2022, pp. 551--558.

\bibitem{bpe_music}
N.~Fradet, N.~Gutowski, F.~Chhel, and J.~Briot, ``Byte pair encoding for symbolic music,'' in \emph{Proceedings of the 2023 Conference on Empirical Methods in Natural Language Processing, {EMNLP} 2023, Singapore, December 6-10, 2023}.\hskip 1em plus 0.5em minus 0.4em\relax Association for Computational Linguistics, 2023, pp. 2001--2020.

\bibitem{FME}
Z.~Guo, J.~Kang, and D.~Herremans, ``A domain-knowledge-inspired music embedding space and a novel attention mechanism for symbolic music modeling,'' in \emph{Thirty-Seventh {AAAI} Conference on Artificial Intelligence, {AAAI} 2023, Washington, DC, USA, February 7-14}, 2023, pp. 5070--5077.

\bibitem{lattner_interval_2018}
S.~Lattner, M.~Grachten, and G.~Widmer, ``Learning transposition-invariant interval features from symbolic music and audio,'' in \emph{Proceedings of the 19th International Society for Music Information Retrieval Conference, {ISMIR} 2018, Paris, France, September 23-27}, 2018.

\bibitem{lattner_interval_2018_predictive}
------, ``A predictive model for music based on learned interval representations,'' in \emph{Proceedings of the 19th International Society for Music Information Retrieval Conference, {ISMIR} 2018, Paris, France, September 23-27}, 2018, pp. 26--33.

\bibitem{Liutkus_rel_attn}
A.~Liutkus, O.~C{\'{\i}}fka, S.~Wu, U.~Simsekli, Y.~Yang, and G.~Richard, ``Relative positional encoding for transformers with linear complexity,'' in \emph{Proceedings of the 38th International Conference on Machine Learning, {ICML} 2021, 18-24 July 2021, Virtual Event}, ser. Proceedings of Machine Learning Research, vol. 139, 2021, pp. 7067--7079.

\bibitem{relative_attention}
P.~Shaw, J.~Uszkoreit, and A.~Vaswani, ``Self-attention with relative position representations,'' in \emph{Proceedings of the 2018 Conference of the North American Chapter of the Association for Computational Linguistics: Human Language Technologies, NAACL-HLT, New Orleans, Louisiana, USA, June 1-6, Volume 2 (Short Papers)}.\hskip 1em plus 0.5em minus 0.4em\relax Association for Computational Linguistics, 2018, pp. 464--468.

\bibitem{rope}
J.~Su, M.~H.~M. Ahmed, Y.~Lu, S.~Pan, W.~Bo, and Y.~Liu, ``Roformer: Enhanced transformer with rotary position embedding,'' \emph{Neurocomputing}, vol. 568, p. 127063, 2024.

\bibitem{guo_musiac}
R.~Guo, I.~Simpson, C.~Kiefer, T.~Magnusson, and D.~Herremans, ``Musiac: An extensible generative framework for music infilling applications with multi-level control,'' in \emph{Artificial Intelligence in Music, Sound, Art and Design - 11th International Conference, EvoMUSART 2022, Held as Part of EvoStar 2022, Madrid, Spain, April 20-22, Proceedings}, ser. Lecture Notes in Computer Science, vol. 13221, 2022, pp. 341--356.

\bibitem{lee_commu}
H.~Lee, T.~Kim, H.~Kang, M.~Ki, H.~Hwang, K.~Park, S.~Han, and S.~J. Kim, ``Commu: Dataset for combinatorial music generation,'' in \emph{Advances in Neural Information Processing Systems 35: Annual Conference on Neural Information Processing Systems 2022, NeurIPS 2022, New Orleans, LA, USA, November 28 - December 9}, 2022.

\bibitem{impact_time_duration}
N.~Fradet, N.~Gutowski, F.~Chhel, and J.~Briot, ``Impact of time and note duration tokenizations on deep learning symbolic music modeling,'' in \emph{Proceedings of the 24th International Society for Music Information Retrieval Conference, {ISMIR} 2023, Milan, Italy, November 5-9}, 2023, pp. 89--97.

\bibitem{figaro}
D.~von R{\"{u}}tte, L.~Biggio, Y.~Kilcher, and T.~Hofmann, ``{FIGARO:} controllable music generation using learned and expert features,'' in \emph{The Eleventh International Conference on Learning Representations, {ICLR} 2023, Kigali, Rwanda, May 1-5}, 2023.

\bibitem{nested_music_transformer}
J.~Ryu, H.~Dong, J.~Jung, and D.~Jeong, ``Nested music transformer: Sequentially decoding compound tokens in symbolic music and audio generation,'' in \emph{Proceedings of the 25th International Society for Music Information Retrieval Conference, {ISMIR} 2024, San Francisco, California, {USA} and Online, November 10-14}, 2024, pp. 588--595.

\bibitem{transformer}
A.~Vaswani, N.~Shazeer, N.~Parmar, J.~Uszkoreit, L.~Jones, A.~N. Gomez, L.~Kaiser, and I.~Polosukhin, ``Attention is all you need,'' in \emph{Advances in Neural Information Processing Systems 30: Annual Conference on Neural Information Processing Systems 2017, December 4-9, 2017, Long Beach, CA, {USA}}, 2017, pp. 5998--6008.

\bibitem{llama}
\BIBentryALTinterwordspacing
H.~Touvron, T.~Lavril, G.~Izacard, X.~Martinet, M.~Lachaux, T.~Lacroix, B.~Rozi{\`{e}}re, N.~Goyal, E.~Hambro, F.~Azhar, A.~Rodriguez, A.~Joulin, E.~Grave, and G.~Lample, ``Llama: Open and efficient foundation language model,'' \emph{CoRR}, vol. abs/2302.13971, 2023. [Online]. Available: \url{https://doi.org/10.48550/arXiv.2302.13971}
\BIBentrySTDinterwordspacing

\bibitem{radford2019language_gpt2}
\BIBentryALTinterwordspacing
A.~Radford, J.~Wu, R.~Child, D.~Luan, D.~Amodei, and I.~Sutskever, ``Language models are unsupervised multitask learners,'' 2019, openAI Blog. [Online]. Available: \url{https://cdn.openai.com/better-language-models/language_models_are_unsupervised_multitask_learners.pdf}
\BIBentrySTDinterwordspacing

\bibitem{transformer-xl}
Z.~Dai, Z.~Yang, Y.~Yang, J.~G. Carbonell, Q.~V. Le, and R.~Salakhutdinov, ``Transformer-xl: Attentive language models beyond a fixed-length context,'' in \emph{Proceedings of the 57th Conference of the Association for Computational Linguistics, {ACL} 2019, Florence, Italy, July 28- August 2, 2019, Volume 1: Long Papers}, 2019, pp. 2978--2988.

\bibitem{linear_transformer}
A.~Katharopoulos, A.~Vyas, N.~Pappas, and F.~Fleuret, ``Transformers are rnns: Fast autoregressive transformers with linear attention,'' in \emph{Proceedings of the 37th International Conference on Machine Learning, {ICML} 2020, 13-18 July 2020, Virtual Event}, ser. Proceedings of Machine Learning Research, vol. 119.\hskip 1em plus 0.5em minus 0.4em\relax {PMLR}, 2020, pp. 5156--5165.

\bibitem{lakhnes}
C.~Donahue, H.~H. Mao, Y.~E. Li, G.~W. Cottrell, and J.~McAuley, ``Lakhnes: Improving multi-instrumental music generation with cross-domain pre-training,'' in \emph{ISMIR}, 2019.

\bibitem{Pasquier_midigpt}
P.~Pasquier, J.~Ens, N.~Fradet, P.~Triana, D.~Rizzotti, J.~Rolland, and M.~Safi, ``{MIDI-GPT:} {A} controllable generative model for computer-assisted multitrack music composition,'' in \emph{AAAI-25, Sponsored by the Association for the Advancement of Artificial Intelligence, February 25 - March 4, 2025, Philadelphia, PA, {USA}}, 2025, pp. 1474--1482.

\bibitem{Lee_gigamidi}
K.~J.~M. Lee, J.~Ens, S.~Adkins, P.~Sarmento, M.~Barthet, and P.~Pasquier, ``The gigamidi dataset with features for expressive music performance detection,'' \emph{Trans. Int. Soc. Music. Inf. Retr.}, vol.~8, no.~1, 2025.

\bibitem{raffel_lakh}
C.~Raffel, ``Learning-based methods for comparing sequences, with applications to audio-to-midi alignment and matching,'' Ph.D. dissertation, Columbia University, {USA}, 2016.

\bibitem{bradshaw2025ariamidi}
L.~Bradshaw and S.~Colton, ``Aria-{MIDI}: A dataset of piano {MIDI} files for symbolic music modeling,'' in \emph{The Thirteenth International Conference on Learning Representations}, 2025.

\bibitem{mustango}
J.~Melechovsk{\'{y}}, Z.~Guo, D.~Ghosal, N.~Majumder, D.~Herremans, and S.~Poria, ``Mustango: Toward controllable text-to-music generation,'' in \emph{Proceedings of the 2024 Conference of the North American Chapter of the Association for Computational Linguistics: Human Language Technologies (Volume 1: Long Papers), {NAACL} 2024, Mexico City, Mexico, June 16-21}, 2024, pp. 8293--8316.

\bibitem{Ainslie_GQA}
J.~Ainslie, J.~Lee{-}Thorp, M.~de~Jong, Y.~Zemlyanskiy, F.~Lebr{\'{o}}n, and S.~Sanghai, ``{GQA:} training generalized multi-query transformer models from multi-head checkpoints,'' in \emph{Proceedings of the 2023 Conference on Empirical Methods in Natural Language Processing, {EMNLP} 2023, Singapore, December 6-10}, 2023, pp. 4895--4901.

\bibitem{mmm}
\BIBentryALTinterwordspacing
J.~Ens and P.~Pasquier, ``{MMM} : Exploring conditional multi-track music generation with the transformer,'' \emph{CoRR}, vol. abs/2008.06048, 2020. [Online]. Available: \url{https://arxiv.org/abs/2008.06048}
\BIBentrySTDinterwordspacing

\bibitem{hu_lora}
E.~J. Hu, Y.~Shen, P.~Wallis, Z.~Allen{-}Zhu, Y.~Li, S.~Wang, L.~Wang, and W.~Chen, ``Lora: Low-rank adaptation of large language models,'' in \emph{The Tenth International Conference on Learning Representations, {ICLR} 2022, Virtual Event, April 25-29}, 2022.

\bibitem{Micikevicius_mixed_precision_training}
P.~Micikevicius, S.~Narang, J.~Alben, G.~F. Diamos, E.~Elsen, D.~Garc{\'{\i}}a, B.~Ginsburg, M.~Houston, O.~Kuchaiev, G.~Venkatesh, and H.~Wu, ``Mixed precision training,'' in \emph{6th International Conference on Learning Representations, {ICLR} 2018, Vancouver, BC, Canada, April 30 - May 3, 2018, Conference Track Proceedings}, 2018.

\bibitem{pijama}
D.~Edwards, S.~Dixon, and E.~Benetos, ``Pijama: Piano jazz with automatic {MIDI} annotations,'' \emph{Trans. Int. Soc. Music. Inf. Retr.}, vol.~6, no.~1, pp. 89--102, 2023.

\bibitem{hung_emopia}
H.~Hung, J.~Ching, S.~Doh, N.~Kim, J.~Nam, and Y.~Yang, ``{EMOPIA:} {A} multi-modal pop piano dataset for emotion recognition and emotion-based music generation,'' in \emph{Proceedings of the 22nd International Society for Music Information Retrieval Conference, {ISMIR} 2021, Online, November 7-12}, 2021, pp. 318--325.

\bibitem{kong_giantpianomidi}
Q.~Kong, B.~Li, J.~Chen, and Y.~Wang, ``Giantmidi-piano: {A} large-scale {MIDI} dataset for classical piano music,'' \emph{Trans. Int. Soc. Music. Inf. Retr.}, vol.~5, no.~1, pp. 87--98, 2022.

\bibitem{ASAP}
F.~Foscarin, A.~McLeod, P.~Rigaux, F.~Jacquemard, and M.~Sakai, ``{ASAP:} a dataset of aligned scores and performances for piano transcription,'' in \emph{Proceedings of the 21th International Society for Music Information Retrieval Conference, {ISMIR} 2020, Montreal, Canada, October 11-16, 2020}, 2020, pp. 534--541.

\bibitem{ATEPP}
H.~Zhang, J.~Tang, S.~R. Rafee, S.~Dixon, G.~Fazekas, and G.~A. Wiggins, ``{ATEPP:} {A} dataset of automatically transcribed expressive piano performance,'' in \emph{Proceedings of the 23rd International Society for Music Information Retrieval Conference, {ISMIR} 2022, Bengaluru, India, December 4-8}, 2022, pp. 446--453.

\bibitem{dadagp}
P.~Sarmento, A.~Kumar, C.~J. Carr, Z.~Zukowski, M.~Barthet, and Y.~Yang, ``Dadagp: {A} dataset of tokenized guitarpro songs for sequence models,'' in \emph{Proceedings of the 22nd International Society for Music Information Retrieval Conference, {ISMIR} 2021, Online, November 7-12}, 2021, pp. 610--617.

\bibitem{filobass}
X.~Riley and S.~Dixon, ``Filobass: {A} dataset and corpus based study of jazz basslines,'' in \emph{Proceedings of the 24th International Society for Music Information Retrieval Conference, {ISMIR} 2023, Milan, Italy, November 5-9}, 2023, pp. 500--507.

\bibitem{filosax}
D.~Foster and S.~Dixon, ``Filosax: {A} dataset of annotated jazz saxophone recordings,'' in \emph{Proceedings of the 22nd International Society for Music Information Retrieval Conference, {ISMIR} 2021, Online, November 7-12}, 2021, pp. 205--212.

\bibitem{GAPS}
X.~Riley, Z.~Guo, A.~C. Edwards, and S.~Dixon, ``{GAPS:} {A} large and diverse classical guitar dataset and benchmark transcription model,'' in \emph{Proceedings of the 25th International Society for Music Information Retrieval Conference, {ISMIR} 2024, San Francisco, California, {USA} and Online, November 10-14}, 2024, pp. 611--617.

\bibitem{groove}
J.~Gillick, A.~Roberts, J.~Engel, D.~Eck, and D.~Bamman, ``Learning to groove with inverse sequence transformations,'' in \emph{International Conference on Machine Learning (ICML)}, 2019.

\bibitem{Guitarset}
Q.~Xi, R.~M. Bittner, J.~Pauwels, X.~Ye, and J.~P. Bello, ``Guitarset: {A} dataset for guitar transcription,'' in \emph{Proceedings of the 19th International Society for Music Information Retrieval Conference, {ISMIR} 2018, Paris, France, September 23-27}, 2018, pp. 453--460.

\bibitem{maestro}
C.~Hawthorne, A.~Stasyuk, A.~Roberts, I.~Simon, C.-Z.~A. Huang, S.~Dieleman, E.~Elsen, J.~Engel, and D.~Eck, ``Enabling factorized piano music modeling and generation with the {MAESTRO} dataset,'' in \emph{International Conference on Learning Representations}, 2019.

\bibitem{MAPS}
V.~Emiya, N.~Bertin, B.~David, and R.~Badeau, ``{MAPS} - a piano database for multipitch estimation and automatic transcription of music,'' INRIA, France, Research Report, Jul. 2010.

\bibitem{metamidi}
J.~Ens and P.~Pasquier, ``Building the metamidi dataset: Linking symbolic and audio musical data,'' in \emph{Proceedings of the 22nd International Society for Music Information Retrieval Conference, {ISMIR} 2021, Online, November 7-12}, 2021, pp. 182--188.

\bibitem{musicnet}
J.~Thickstun, Z.~Harchaoui, and S.~M. Kakade, ``Learning features of music from scratch,'' in \emph{5th International Conference on Learning Representations, {ICLR} 2017, Toulon, France, April 24-26, 2017, Conference Track Proceedings}, 2017.

\bibitem{supra}
Z.~Shi, C.~S. Sapp, K.~Arul, J.~McBride, and J.~O. Smith, ``Supra: Digitizing the stanford university piano roll archive.'' in \emph{Proceedings of the 20th International Society for Music Information Retrieval}, Delft, The Netherlands, 2019, pp. 517--523.

\bibitem{urmp}
B.~Li, X.~Liu, K.~Dinesh, Z.~Duan, and G.~Sharma, ``Creating a multitrack classical music performance dataset for multimodal music analysis: Challenges, insights, and applications,'' \emph{IEEE Transactions on Multimedia}, vol.~21, no.~2, pp. 522--535, 2019.

\bibitem{weimarjazz}
M.~Pfleiderer, K.~Frieler, J.~Abe{\ss}er, W.-G. Zaddach, and B.~Burkhart, Eds., \emph{{I}nside the {J}azzomat - {N}ew {P}erspectives for {J}azz {R}esearch}.\hskip 1em plus 0.5em minus 0.4em\relax Schott Campus, 2017.

\end{thebibliography}

\begin{appendices}
\section{Training and Test Data used for Pretraining} \label{sec:training_datasets}
A summary of the training data used for pretaining can be found in Table \ref{tab:training_datasets}. During pretraining, rather than splitting test data from the listed datasets, we constructed a private test set consisting of real human performance MIDI to minimize overlap between the train and test sets. 
\begin{table}[ht]
  \caption{Summary of the Training Datasets.}
  \label{tab:training_datasets}
  \centering
      \resizebox{\columnwidth}{!}{%
  \begin{tabular}{lcccc}
    \toprule
    Dataset & Description & Duration (Hours) & Number of Tokens \tablefootnote{Each event $x$ consists of 6 tokens: onset, duration, octave, pitch class, instrument, velocity.} & License\\
    \midrule
    AriaMIDI \cite{bradshaw2025ariamidi} & Piano & 57380.32  & 8.40B &  CC-BY-NC-SA 4.0\\
    ASAP~\cite{ASAP} & Piano & 110.57  & 25.68M & CC BY-NC-SA 4.0\\
    ATEPP~\cite{ATEPP} & Piano & 997.11  & 194.82M & CC BY 4.0\\
    DadaGP~\cite{dadagp} & Guitar, Multitrack & 1124.29  & 435.00M & Research-only \\
    Doug McKenzie\tablefootnote{\url{bushgrafts.com}} & Jazz Combo & 22.77  & 5.74M & - \\
    FiloBass \cite{filobass} & Bass & 4.94  & 315.36K  & CC BY 4.0\\
    FiloSax \cite{filosax} & Saxophone & 24.55  & 1.74M & Non-commercial, research-only\tablefootnote{\href{https://dave-foster.github.io/filosax/}{https://dave-foster.github.io/filosax/}} \\
    GAPS \cite{GAPS} & Classical Guitar & 15.69  & 1.61M & CC BY-NC-SA 4.0 \\
    Groove \cite{groove} & Drum & 13.54  & 2.68M & CC BY 4.0 \\
    GuitarSet \cite{Guitarset} & Guitar & 3.03  & 374.88K &CC BY 4.0 \\
    Maestro \cite{maestro} & Piano & 197.33  & 42.24M & CC BY-NC-SA 4.0\\
    MAPS \cite{MAPS} & Piano & 31.69  & 4.61M & CC BY-NC-SA 2.0\\
    MetaMIDI~\cite{metamidi} & Multitrack & 18189.58  & 7.26B & Research-only \\
    MusicNet~\cite{musicnet} & Classical & 33.4  & 6.36M & CC BY 4.0 \\
    PiJAMA~\cite{pijama} & Jazz Piano & 217.1  & 40.62M & CC BY-NC 4.0\\
    SUPRA~\cite{supra} & Piano & 48.45  & 10.68M & CC BY-NC-SA 4.0 \\
    SymphonyNet~\cite{symphonynet} & Classical & 3135.99  & 1.61B & -\\
    URMP~\cite{urmp} & Classical & 1.35  & 177.66K & - \\
    Weimar Jazz~\cite{weimarjazz} & Jazz & 12.86  & 1.19M & - \\
    \midrule
    \textbf{Total} & - & 81.58K  & 18.06B \\
    \bottomrule
  \end{tabular}}
\end{table}

\section{Limitations} \label{sec:limitation}

As shown in Table \ref{tab:training_datasets}, our training data mainly consists of Western music, which is likely to introduce biases. Additionally, our model is limited to processing only note-level MIDI information, while other information present in the MIDI file, such as metadata, system messages, controller data (e.g., CC values used for modulation, volume, and sustain parameters), is excluded.

\section{Ethical Considerations, Envirommental Implications and Other Broader Impacts}\label{sec:broader_impacts}
Generative models pose both legal and ethical challenges to our society. As both researchers and musicians, our goal is to make scientific progress—which relies on high-quality data—while at the same time upholding fair use of training data and respect for the rights of license holders. Hence, we choose to use datasets which are linked with previous publications, and clearly list the data used for pretraining in Table \ref{tab:training_datasets}. However, we acknowledge that some of these datasets involve web scraping or music transcription, which can introduce ambiguity regarding licensing. We encourage future researchers to disclose the details of their training data especially for large-scale pretraining. This would not only promote fair use of training data and protect the rights of license owners, but also facilitate more meaningful comparisons between models, since currently we are unclear about whether the model differences come from the training data or the model architecture. 

Training foundation models can have significant environmental impacts. Due to our efficient data encoding and model architecture, we substantially reduce computational costs compared to other similar work. For instance, training Clamp2 \cite{wu_clamp2} and M3 \cite{wu_clamp2} required 8 NVIDIA H800 GPUs and took approximately 33.3 days; Training Mupt (1.97B) \cite{qu2025mupt} used 32 A100 GPUs over 11.3 days. Meanwhile, training \modelname{} (M) only required 2 A100 GPUs for 15 days. We encourage future researchers to innovate in model design and optimization, as such efforts can contribute to improved training efficiency and environmental sustainability.

Next, we discuss the economic implications of releasing \modelname{}. Since \modelname{} has been pretrained on a large and diverse collection of MIDI data, the pretrained model contains a ``compression'' of human's collective musical knowledge, that may improve productivity and augment human capabilities, or may cause job displacements in the music labor markets. The ethical statements in both Anticipatory Music Transformer \cite{anticipatory_music_transformer} and GAPS \cite{GAPS} express similar concerns. Although the overall impact on the music labor market remains unclear, we support the view stated in \cite{anticipatory_music_transformer} that researchers should create systems that enhance productivity and augment human intelligence, rather than fully automate the creative process. 

Another potential impact of \modelname{} lies in the transferability of its architecture to other domains. Although our primary focus is on computer music, the underlying model architecture may be applicable to a broader range of tasks that involve high-dimensional data (e.g., robotics). While this opens up opportunities in other fields, it also raises ethical concerns regarding broader societal impacts. We are unable to provide a clear picture of the exact consequences it may have, but we acknowledge that our model could have other far-reaching effects.


\section{Model Configurations} \label{sec:model_configs}
The configurations of our pretrained models \modelname{} (S) and \modelname{} (M) are shown in Table \ref{tab:model_size}.
\begin{table}[ht]
  \caption{Model configurations}
  \label{tab:model_size}
  \centering
  \begin{tabular}{lcc}
    \toprule
    & \modelname{} (S) & \modelname{} (M) \\
    \midrule
    Model Params & \smallmodelsize{} & \largemodelsize{} \\
    Hidden Size & 1536 & 1920 \\
    Intermediate Size & 5376 & 6720 \\
    Heads (Q) & 12 & 12 \\
    Heads (K, V) & 6 & 6 \\
    Layers (Attn) & 9 & 15 \\
    Layers (GRU) & 2 & 4 \\
    Base value of $\theta_o$ & 199999& 199999 \\
    Base value of $\theta_d$ & 1031& 1031 \\
    Base value of $\theta_{oct}$ & 19& 19 \\
    Base value of $\theta_{p}$ & 20& 20 \\
    Base value of $\theta_{v}$ & 131& 131 \\
    GRU Hidden Size & 1024 & 1536 \\
    GRU Output Size & 2341 & 8487 \\
    Training Data & LakhMIDI \cite{raffel_lakh} & Table~\ref{tab:training_datasets} \\
    \bottomrule
  \end{tabular}
\end{table}
\section{Data Preprocessing}\label{sec:data_preprocessing}
Although FME \cite{FME} is based on a continuous sinusoidal embedding function which allows it to extrapolate or interpolate to unseen inputs (e.g., large onset times, microtonal pitches), we model the categorical distribution of each sub-event in $\tilde{x}$ during pretraining, which requires quantization. Hence, for each MIDI file, we first quantize time into 10ms intervals, which allows us to represent onsets and timeshifts as discrete values while maintaining performance timing. We set the maximum allowed timeshifts to 10240ms for \modelname{} (S) and 40960ms for \modelname{} (M). Similarly, the maximum note durations are limited to 10240ms and 40960ms for \modelname{} (S) and \modelname{} (M), respectively. MIDI files exceeding these limits are discarded. We then process each MIDI file into 2 sequences of tuples in the format of $x$ and $\tilde{x}$ following the notation introduced in Section \ref{sec:method}. The statistics of the training data after preprocessing are summarized in Table~\ref{tab:training_datasets}.


At the GRU, the output size matches the size of the token dictionary, which includes:
\begin{itemize}
    \item 1 \texttt{<sos\_gru>} token,
    \item 4,097 time shift tokens for \modelname{} (M) or 1,024 for \modelname{} (S),
    \item 4,097 duration tokens for \modelname{} (M) or 1,024 for \modelname{} (S),
    \item 11 octave tokens,
    \item 12 pitch class tokens,
    \item 129 instrument tokens,
    \item 128 velocity tokens,
    \item 1 \texttt{<sos\_$\tilde{x}$>} and 1 \texttt{<eos\_$\tilde{x}$>} token for each attribute in $\tilde{x}$ (12 in total).
\end{itemize}

After pretraining, the quantization constraints and the filtering of long durations or timeshifts can be omitted depending on finetuning tasks, since FME \cite{FME} can process unseen inputs. For instance, in the music classification tasks, we encountered some cases where some note durations exceed the maximum duration allowed during pretraining. In this case, since we do not require a GRU sub-decoder in the finetuning architecture (see Section \ref{sec:method_finetune}) to predict the next event, we can safely keep these data during finetuning. 

\section{Encoding relative positional information in high dimensions with \newattentionshort{}} \label{sec:MRA_derivation}

Following the notation introduced in Section \ref{sec:method}, we derive how relative position information can be encoded in \newattentionshort{} in higher dimensions. We start from a 1-D case with RoPE where a query $\mathbf{Q}^m =  [\mathbf{Q}_1^m, \mathbf{Q}_2^m, \dots, \mathbf{Q}_H^m]$ and a key vector $\mathbf{K}^n = [\mathbf{K}_1^n, \mathbf{K}_2^n, \dots, \mathbf{K}_H^n]$ are located at positon $m$ and $n$ respectively, both with $H$ heads. For simplicity, we ignore the batch dimension and both $\mathbf{Q}^m$ and $\mathbf{K}^n$ have shape $(H, 1, d_h)$. In RoPE, the query and key vectors are first rotated based on their position: $\mathbf{Q}^m e^{i m \theta} = [\mathbf{Q}_1^me^{i m \theta}, \dots, \mathbf{Q}_H^me^{i m \theta}]$, $\mathbf{K}^n e^{i n \theta} = [\mathbf{K}_1^ne^{i n \theta}, \dots, \mathbf{K}_H^ne^{i n \theta}]$. The unnormalized attention score is calculated in Equation \ref{eq:rope_qk_decomposed}. We ignore the normalization term for simplicity here. 

\begin{equation}\label{eq:rope_qk_decomposed}
\langle \mathbf{Q}^m e^{i m \theta}, \mathbf{K}^n e^{i n \theta} \rangle  = \Re\left[\mathbf{Q}^m \mathbf{K}^{n^*} e^{i (m-n) \theta}\right] =\left[\Re\left[\mathbf{Q}_1^m \mathbf{K}_1^{n^*} e^{i (m-n) \theta}\right], \dots, \Re\left[\mathbf{Q}_H^m \mathbf{K}_H^{n^*} e^{i (m-n) \theta}\right]\right]
\end{equation}


We define the unnormalized attention value for head $h$ with relative distance \( m - n \) as
\[
A_{h, \theta}^{m - n} := \Re\left[\mathbf{Q}_h^m (\mathbf{K}_h^n)^* e^{i (m - n) \theta} \right].
\]
Hence, the unnormalized attention score tensor with shape $(H, 1, 1)$ in Equation \ref{eq:rope_qk_decomposed} can be abbreviated as:

\[
\langle \mathbf{Q}^m e^{i m \theta}, \mathbf{K}^n e^{i n \theta} \rangle  = \left[A_{1, \theta}^{m - n}, \dots,A_{H, \theta}^{m - n}\right].
\]

Next, we extend the relative attention calculation to the 2-D case using \newattentionshort{}, where the query and key vector $\mathbf{Q}^{(a_q, b_q)} =  [\mathbf{Q}_1^{(a_q, b_q)}, \mathbf{Q}_2^{(a_q, b_q)}, \dots, \mathbf{Q}_H^{(a_q, b_q)}]$ and $\mathbf{K}^{(a_k, b_k)} = [\mathbf{K}_1^{(a_k, b_k)}, \mathbf{K}_2^{(a_k, b_k)}, \dots, \mathbf{K}_H^{(a_k, b_k)}]$ with $H$ heads are located at $(a_q, b_q)$ and $(a_k, b_k)$ respectively. Following the formulation of \newattentionshort{} in Section \ref{sec:method}, we partition $H$ heads into two groups with the first $s$ heads assigned to the first group $\mathcal{G}_1 = \{1, \dots, s\}$, and the rest to the second group $\mathcal{G}_2 = \{s+1, \dots, H\}$, where $1<s<H-1$. The heads in $\mathcal{G}_1$ and $\mathcal{G}_2$ are rotated based on the position decomposition along the $a$ and $b$ axes respectively, shown in Equations \ref{eq:rope_q_high_dimension} and \ref{eq:rope_k_high_dimension}.

\begin{equation} \label{eq:rope_q_high_dimension}
    f_q(\mathbf{x}^{(a_q, b_q)}, a_q, b_q) := [\mathbf{Q}_1^{a_q}e^{i a_q \theta_a}, \dots, \mathbf{Q}_s^{a_q}e^{i a_q \theta_a}, \mathbf{Q}_{s+1}^{b_q}e^{i b_q \theta_b}, \dots, \mathbf{Q}_{H}^{b_q}e^{i b_q \theta_b}]
\end{equation}
\begin{equation}\label{eq:rope_k_high_dimension}
    f_k(\mathbf{x}^{(a_k, b_k)}, a_k, b_k) := [\mathbf{K}_1^{a_k}e^{i a_k \theta_a}, \dots, \mathbf{K}_s^{a_k}e^{i a_k \theta_a}, \mathbf{K}_{s+1}^{b_k}e^{i b_k \theta_b}, \dots, \mathbf{K}_{H}^{b_k}e^{i b_k \theta_b}]
\end{equation}

Similarly, the unnormalized attention score tensor with shape $(H, 1, 1)$ is calculated in Equation \ref{eq:rope_qk_decomposed_2d}. As a result, the first $s$ heads encode relative position along the first axis, and the remaining heads encode relative position along the second axis, while still retaining the original attention formulation.

\begin{equation}\label{eq:rope_qk_decomposed_2d}
\langle f_q(\mathbf{x}^{(a_q, b_q)}, a_q, b_q), f_k(\mathbf{x}^{(a_k, b_k)}, a_k, b_k) \rangle  = \left[A_{1, \theta_a}^{a_q - a_k},\dots, A_{s, \theta_a}^{a_q - a_k}, A_{s+1, \theta_b}^{b_q - b_k},\dots, A_{H, \theta_b}^{b_q - b_k}\right]
\end{equation}

To extend this to a multidimensional case, we partition the attention heads into $G$ groups and rotate each group of heads based on their assigned position decomposition along each intrinsic dimension. In \modelname{}, each event has 6 sub-tokens and is located in a 5-D space. We equally divide the total number of heads into 6 groups in all query and key vectors and assign the 5-D position to all groups. The extra group is associated with the instrument attribute, which is categorical and does not have a clear representation in the continuous positional space. We hence assign its position as the onset value, which indicates the time at which the instrument is played.
\section{\newattentionshort{} variant} \label{sec:MRA_variant}

The details of the \newattentionshort{} variant used in the ablation study in Section \ref{sec:experiment_setting} are described as follows: Instead of assigning different groups of heads with their corresponding position decomposition, in this variant, we rotate all attention heads using the sum of the scalar values from the position decomposition. More specifically, the rotation is applied as: $\mathbf{Q}^{(a_q, b_q)}e^{i a_q \theta_a}e^{i b_q \theta_b}$ and $\mathbf{K}^{(a_k, b_k)}e^{i a_k \theta_a}e^{i b_k \theta_b}$ for all heads, following the notation introduced in Appendix \ref{sec:MRA_derivation} and Equations \ref{eq:rope_q} and \ref{eq:rope_k} in Section \ref{sec:method}. The unnormalized attention score is then given by Equation \ref{eq:rope_qk_decomposed_2d_variant} below. The \newattentionshort{} variant also encodes relative positional information in higher dimensions.

\begin{equation}\label{eq:rope_qk_decomposed_2d_variant}
\langle \mathbf{Q}^{(a_q, b_q)}e^{i a_q \theta_a}e^{i b_q \theta_b}, \mathbf{K}^{(a_k, b_k)}e^{i a_k \theta_a}e^{i b_k \theta_b} \rangle  = \Re\left[\mathbf{Q}^{(a_q, b_q)} \mathbf{K}^{(a_k, b_k)^*} e^{i (a_q-a_k) \theta_a}e^{i (b_q-b_k) \theta_b}\right]
\end{equation}

\section{Finetuning Settings} \label{sec:finetuning_settings}

\subsection{Music Classification}


We describe how we gather training segments for our music classification tasks. For each dataset, we use a sliding window with a fixed window length with an overlap ratio of 0.25 to gather training segments during preprocessing. The window length is empirically set to half the mean sequence length of the dataset. If the remaining tokens are fewer than the sliding window size, we pad them to match the window size during finetuning. The extracted training segments are prepended with a \texttt{<sos>} token and appended with both a \texttt{<eos>} token and a \texttt{<cls>} token, following the finetuning architecture described in Section~\ref{sec:method_finetune}. Hence, the finetuning sequence length is always equal to the sliding window length plus 3. We finetune the pretrained \modelname{} using LoRA \cite{hu_lora} and set the rank ($r$) to 8 and dropout to 0.05. The full set of hyperparameters used during finetuning is listed in Table \ref{tab:downstream_classification_param}. To perform clip-wise classification during inference, instead of applying a sliding window to the segment the test data, we directly prepend and append the \texttt{<sos>}, \texttt{<eos>}, and \texttt{<cls>} tokens to each data and perform classification.

\begin{table}
  \caption{Hyperparameters used for classification across different datasets.}
  \label{tab:downstream_classification_param}
  \centering
  \begin{tabular}{lcccc}
    \toprule
    & PiJAMA30 & Pianist8 & Emopia & GPM30 \\
    \midrule
    Sliding window length & 1200 & 900 & 130 & 1200 \\
    Finetuning sequence length & 1203 & 903 & 133 & 1203 \\
    LoRA target modules & q, k, v, o & q, k, v, o & q, k & q, k, v, o \\
    Learning rate & 3.0e-4 & 1.6e-4 & 2.0e-4 & 2.4e-4 \\
    \bottomrule
  \end{tabular}
\end{table}

\subsection{Preliminary Experiment for Music Classification}
We conduct a preliminary experiment to validate the effectiveness of the finetuning architecture for player classification and compare our model against a pretrained and finetuned CRNN model \cite{pijama} using the PiJAMA30 dataset \cite{pijama}. To ensure a fair comparison, we perform clip-wise classification following the experimental setup described in \cite{pijama}. During training, we use an overlapping window of 15 seconds with an overlap ratio of 0.125 to extract training segments. After the model is trained, we use a non-overlapping sliding window of 15 seconds to segment the test data and perform classification subsequently. The results in Table \ref{tab:preliminary_classification_test} show that our model outperforms the CRNN baseline and verifies the effectiveness of our finetuning architecture.

\begin{table}
  \caption{PiJAMA30 classification results on track-level and song-level splits using \modelname{} and a pretrained and finetuned CRNN.}
  \label{tab:preliminary_classification_test}
  \centering

  \begin{tabular}{lcc}
    \toprule
    Model     & \multicolumn{1}{c}{PiJAMA Track Split} & \multicolumn{1}{c}{PiJAMA Song Split} \\
    \cmidrule(r){2-2} \cmidrule(r){3-3}
              & Acc & Acc \\
    \midrule
    CRNN \cite{pijama}       & 0.556 & 0.502 \\
    \modelname{}   & \textbf{0.672} & \textbf{0.563} \\
    \bottomrule
  \end{tabular}
\end{table}

\subsection{Construction of the GPM30 Dataset for Composer Classification}

\begin{figure}
    \centering
    \includegraphics[width=0.85\linewidth]{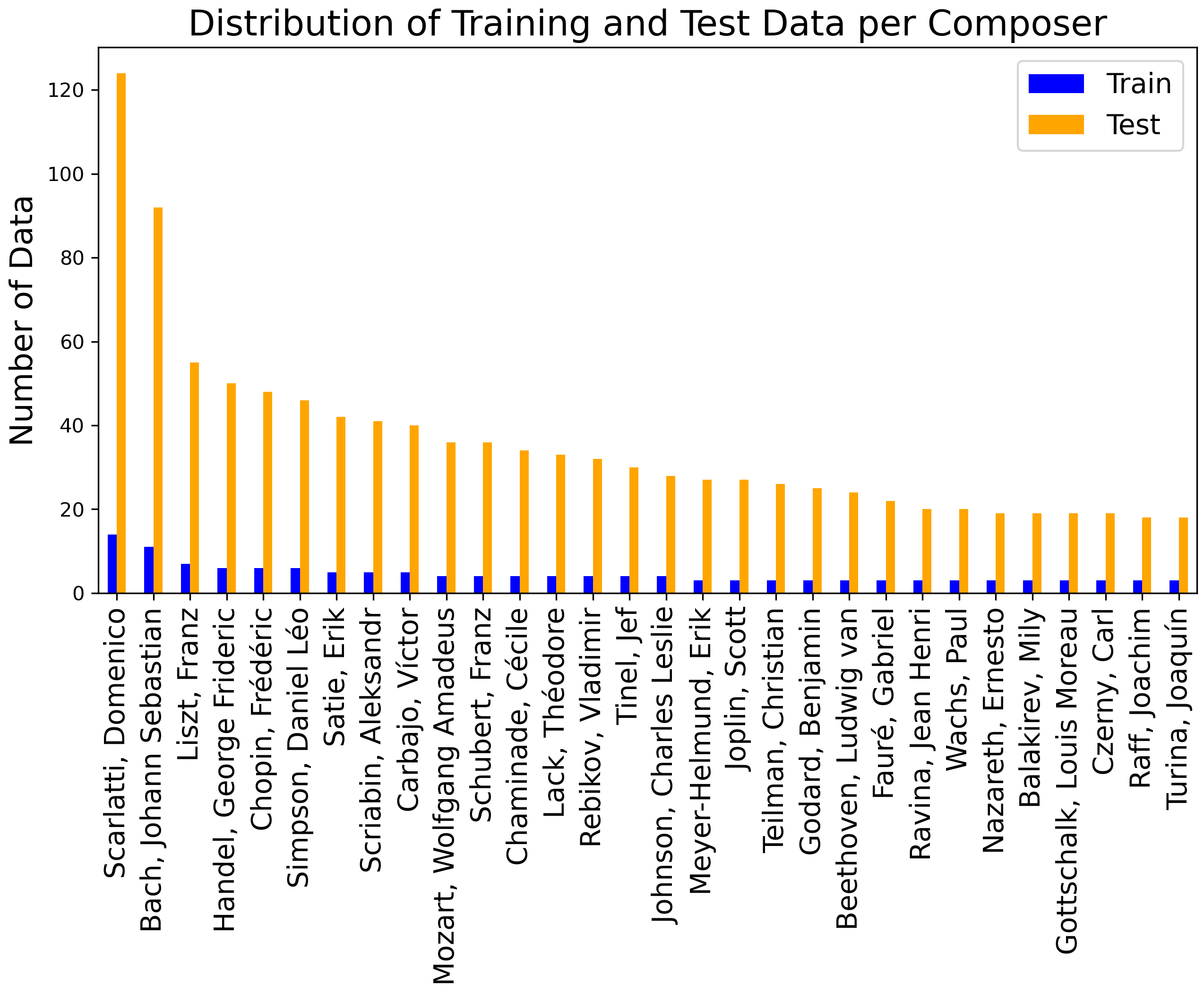}
    \caption{Distribution of the GPM30 dataset.}
    \label{fig:gpm30_distribution}
\end{figure}

Following \cite{pijama}, we construct a smaller subset of the Giant Piano MIDI (GPM) dataset \cite{kong_giantpianomidi} for our composer classification task. We first filter the GPM dataset and only keep the data with a sequence length shorter than 4096. We then sort the filtered dataset by the number of pieces per composer and select the top 30 composers. For each composer, 90\% of the data is randomly sampled for training, and the remaining 10\% is used for testing. This results in a dataset of 1205 pieces, with the distribution across composers shown in Figure \ref{fig:gpm30_distribution}.

\subsection{Conditional Music Generation and Music Infilling}
To increase training efficiency in the conditional music generation task, we normalize the tempo to 120 BPM for all data, since BPM can be easily adjusted during post-processing. However, we still include BPM as a metadata condition since it might influence note choices. During the initial inspection, we found that the model is unable to end a piece properly using the data concatenation method used in pretraining (see Section \ref{sec:pretraining}). This is because the piece of music might be split across two input sequences and the \texttt{<eos>} token only appears in the second sequence. We therefore pad the input sequence to a uniform length of 848, the maximum length under our setting, ensuring that each data point ends with an \texttt{<eos>} token.  Similarly, we use LoRA \cite{hu_lora} for finetuning and set the rank ($r$) to 8 and dropout to 0.05 for LoRA.

\section{Listening Test}\label{sec:listening_test_supplementary}

To conduct the listening test, we first submitted an application to our institution's ethics committee and provided a detailed description of our study, our target participants, a draft questionnaire (distributed via Google Forms), and eventually obtained ethics approval.

\begin{figure}
    \centering
    \includegraphics[width=0.9\linewidth]{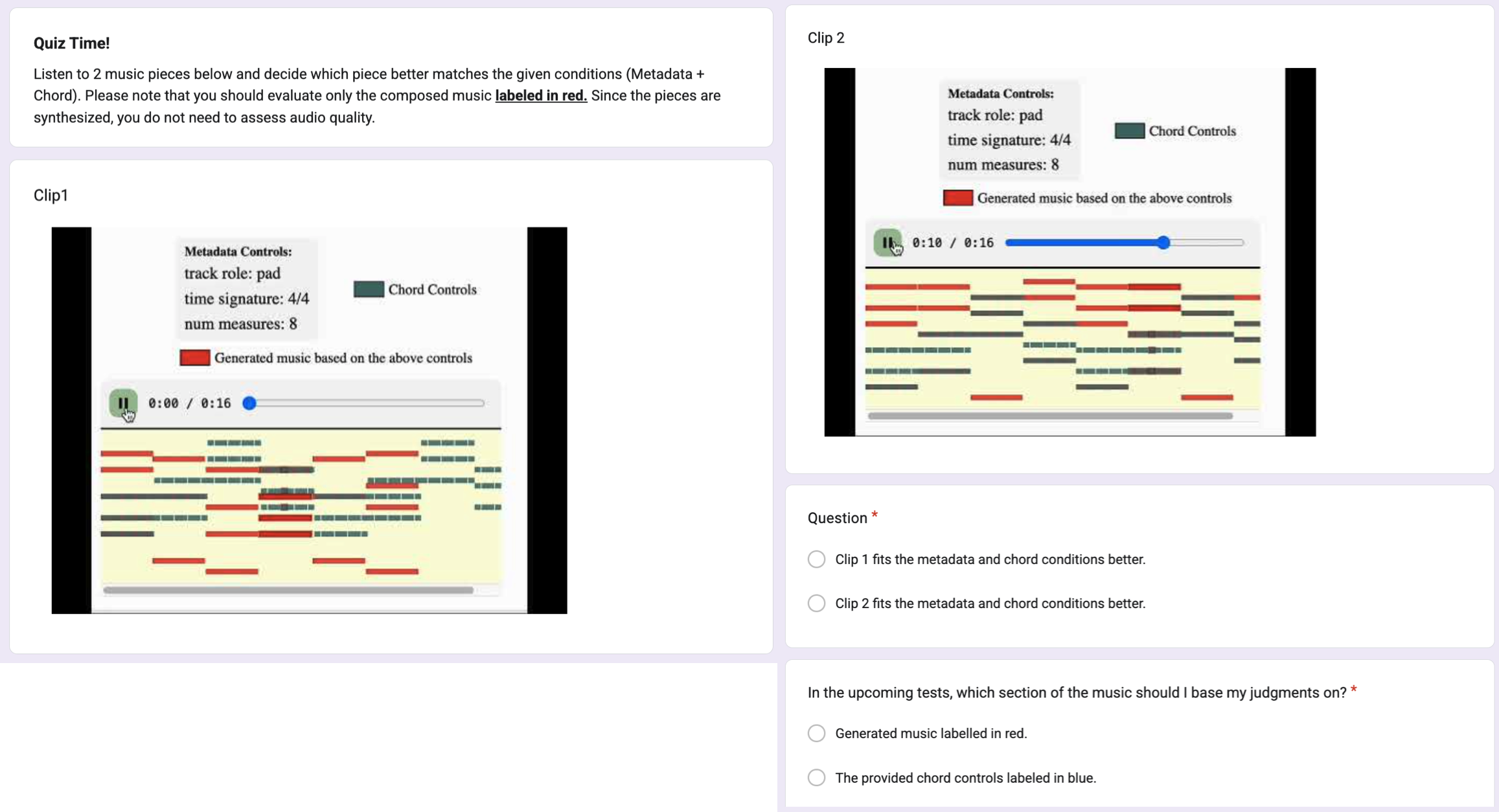}
    \caption{Two questions designed to filter for qualified participants during the listening test.}
    \label{fig:listening_test_quality_control}
\end{figure}

\begin{figure}
    \centering
    \includegraphics[width=0.85\linewidth]{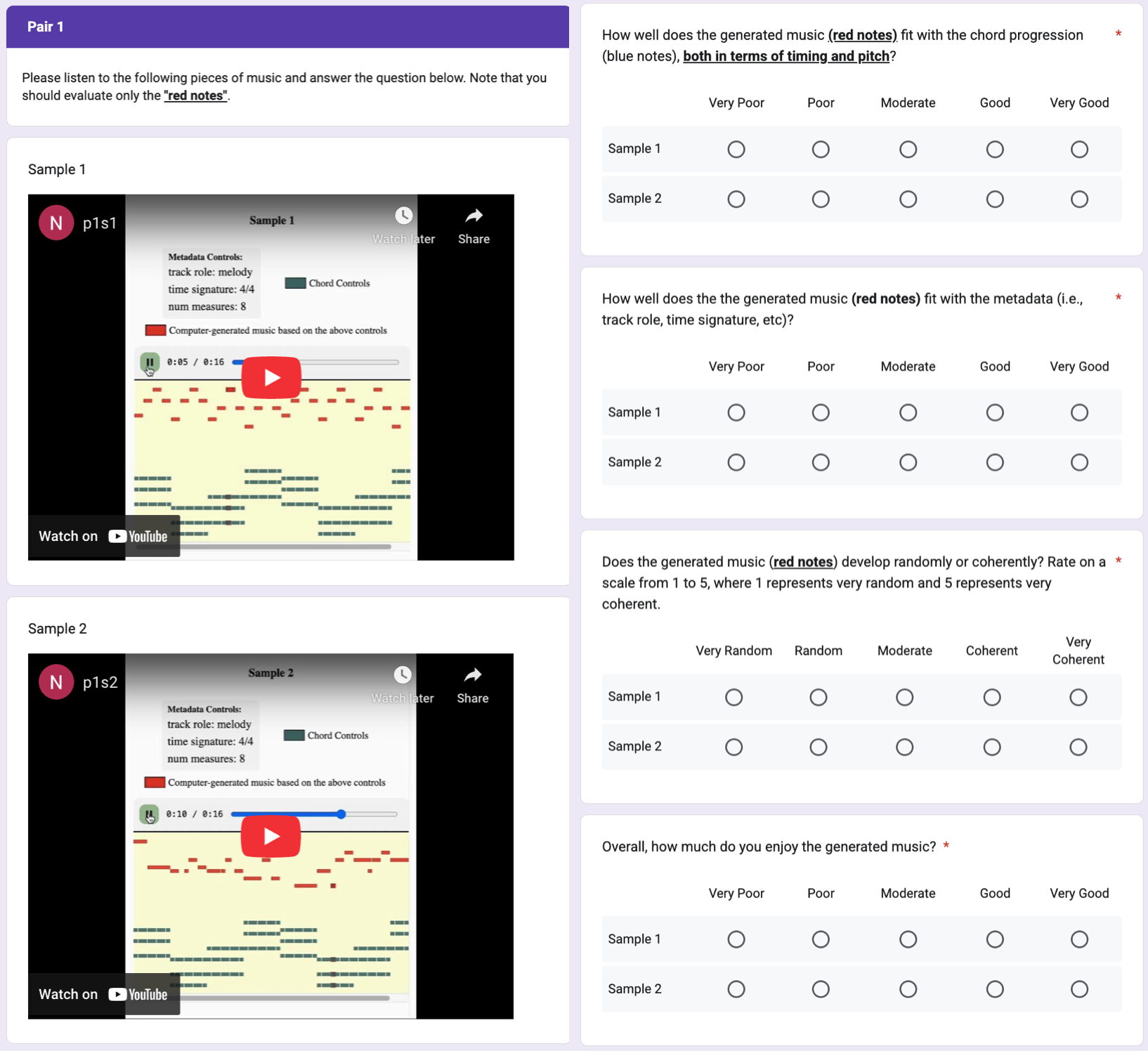}
    \caption{Listening test interface.}
    \label{fig:listening_test_questions}
\end{figure}

We provide an introduction to the participants, including the purpose of the user study, details of the interface, and the types of conditions used. We then ask users to answer two questions designed to filter for qualified participants, as shown in Figure \ref{fig:listening_test_quality_control}. Finally, we ask the participants to rate outputs from the two systems. Because the generated music is conditioned on a velocity range, the volumes of the audio rendered from MIDI vary accordingly. Since this metric can be measured objectively, we normalize the volume across all generated samples to ensure fair comparisons.  A screenshot of the questionnaire is provided in Figure \ref{fig:listening_test_questions}.

\end{appendices}

\end{document}